\documentclass[10pt]{article}
\usepackage[top=1.0in,right=1.0in,left=1.0in,bottom=1.75in]{geometry}
\usepackage{amssymb,amsbsy}
\usepackage{amsmath}
\usepackage{amsthm}
\usepackage{graphics,graphicx}
\usepackage[T1]{fontenc}
\usepackage{color}
\usepackage{subfigure}
\usepackage[affil-it,auth-sc]{authblk}

\newcommand{\beq}{\begin{equation}}
\newcommand{\eeq}{\end{equation}}

\newcommand{\beqar}{\begin{eqnarray}}
\newcommand{\eeqar}{\end{eqnarray}}
\newcommand{\bit}{\begin{itemize}}
\newcommand{\eit}{\end{itemize}}
\newcommand{\benum}{\begin{enumerate}}
\newcommand{\eenum}{\end{enumerate}}
\newcommand{\barr}{\begin{array}}
\newcommand{\earr}{\end{array}}
\def\ds{\displaystyle}

\newcommand{\modIII}{\text{III}}

\def\XXint#1#2#3{{\setbox0=\hbox{$#1{#2#3}{\int}$}
   \vcenter{\hbox{$#2#3$}}\kern-.5\wd0}}

\def\b0{\mbox{\boldmath $0$}}

\def\bn{\mbox{\boldmath $n$}}

\def\bp{\mbox{\boldmath $p$}}
\def\bq{\mbox{\boldmath $q$}}

\def\bt{\mbox{\boldmath $t$}}

\def\bx{\mbox{\boldmath $x$}}

\newcommand{\bsigma}{\mbox{\boldmath $\sigma$}}

\newcommand{\bepsilon}{\mbox{\boldmath $\epsilon$}}

\newcommand{\btau}{\mbox{\boldmath $\tau$}}

\newcommand{\bvarphi}{\mbox{\boldmath$\varphi$}}

\newcommand{\bchi}{\mbox{\boldmath $\chi$}}
\newcommand{\bmu}{\mbox{\boldmath $\mu$}}

\def\f0{\ensuremath{\mathbb{O}}}

\newcommand{\mL}{\ensuremath{\mathcal{L}}}

\def\Im{\mathop{\mathrm{Im}}}

\newcommand{\Reals}{\ensuremath{\mathbb{R}}}

\def\EFM{{\it Eng.\ Fract.\ Mech.}\ }
\def\EJMA{{\it Eur.\ J.\ Mech. A-Solid.}\ }

\def\IJES{{\it Int.\ J.\ Eng.\ Sci.}\ }
\def\IJF{{\it Int.\ J.\ Fracture}\ }

\def\IJSS{{\it Int.\ J.\ Solids Struct.}\ }

\def\JMPS{{\it J.\ Mech.\ Phys.\ Solids}\ }

\title{Steady-state propagation of a Mode III crack in couple stress elastic materials}

\author[1]{G. Mishuris}
\author[1]{A. Piccolroaz\footnote{Corresponding author: e-mail: roaz@ing.unitn.it; phone: +39\,0461\,282583.}}
\author[2]{E. Radi}

\affil[1]{Institute of Mathematical and Physical Sciences, Aberystwyth University, Wales, U.K.}
\affil[1]{Dipartimento di Scienze e Metodi dell'Ingegneria, Universit\`a di Modena e Reggio Emilia, Italy}

\begin{document}

\maketitle

\begin{abstract}
\noindent
This paper is concerned with the problem of a semi-infinite crack steadily propagating in an elastic solid with microstructures subject to antiplane loading
applied on the crack surfaces. The loading is moving with the same constant velocity as that of the crack tip. We assume subsonic regime, that is the crack velocity 
is smaller than the shear wave velocity. The material behaviour is described by the indeterminate theory of couple stress elasticity developed by Koiter. This
constitutive model includes the characteristic lengths in bending and torsion and thus it is able to account for the underlying microstructure of the material
as well as for the strong size effects arising at small scales and observed when the representative scale of the deformation field becomes comparable with the
length scale of the microstructure, such as the grain size in a polycrystalline or granular aggregate.

The present analysis confirms and extends earlier results on the static case by including the effects of crack velocity and rotational inertia. By adopting the
criterion of maximum total shear stress, we discuss the effects of microstructural parameters on the stability of crack propagation.

\end{abstract}


% \newpage

% \tableofcontents
%
% \newpage

\section{Introduction}
\label{sec1}

In view of the discontinuous nature of many engineering materials, it is well known that the classical theory of elasticity is not adequate to accurately describe the stress and 
displacement fields at small scales (Morozov, 1984). The discrepancy between the classical theoretical predictions and experimental results is found more pronounced for materials 
with a coarse-grain structure (Fleck et al., 1994). The mechanical behaviour of most materials with microstructure, like composites, cellular materials, foams, masonry, bone tissues, 
glassy and semicrystalline polymers, is strongly influenced by the microstructural characteristic lengths, especially in the presence of large stress (or strain) gradients 
(Lakes, 1986; Lakes, 1995). These findings stimulated the development of generalized theories of continuum mechanics such as micropolar elasticity (Cosserat, 1909), indeterminate 
couple stress elasticity (Koiter, 1964) and more recently strain gradient theories (Aifantis, 2011; Fleck \& Hutchinson, 2001). An extensive review of various formats of gradient 
elasticity and their performance in static and dynamic applications can be found in Askes \& Aifantis (2011).

Because of the characteristic lengths coming into play, the generalized theories predict dispersive wave propagation in microstructured media (Nowacki, 1985). Also the generalized 
theories predict different asymptotic behaviour for some components of the deformation and stress fields near singular points, compared to the classical elasticity theory 
(Morozov, 1984). Correspondingly, the notion of energy release rate and J-integral needs to be generalized (Lubarda and Markenscoff, 2000; Piccolroaz et al., 2012). Interestingly, 
the relation between stress intensity factors for classic and CS elasticity discussed in detail in Radi (2008) by means of the Wiener--Hopf technique was also found by Mozorov \& 
Nazarov (1980) and by Nazarov \& Semenov (1980). The fact that microrotaions are bounded (and non-zero) at the crack tip has led to the formulation of new fracture criteria 
(Morozov, 1984). 

Due to the complexity of the equations of motion provided by the couple stress elastic theory with rotational inertia, only few crack propagation problems have been considered in 
the literature, most of them have been solved numerically and very few closed form solution have been worked out. In particular, Parhi \& Das (1970) used the couple stress theory 
to investigate the stress distribution in a half-plane due to a moving line load on the surface. They assumed that the load is moving slower than both the longitudinal and transverse 
wave speeds of the elastic media (subsonic case) and found that the couple stress effect depends on the speed of the moving load, the couple-stress parameter, and the Poisson's ratio.
Han et al. (1990) investigated dynamic propagation of a finite crack under mode-I loading in a micropolar elastic solid. 
Since the Wiener--Hopf technique can not be easily applied to finite crack problem, these authors solved numerically a pair of two-dimensional singular integral equations obtained by 
using an integral transform method. They provided solutions for dynamic stress intensity and couple stress intensity factors by using the obtained values of the strengths of the square 
root singularities in macrorotation and the gradient of microrotation at the crack tips.
Itou (1981) considered the steady-state propagation in plane strain condition for the case of vanishing rotational inertia and only one characteristic length, analysing the 
influence of the crack tip speed on the asymptotics of the stress field. 
However, as already pointed out by Zhang et al. (1998) and Radi (2008) for the static case, a simple asymptotic 
characterization of the crack tip fields is not sufficient to analyse the fracture behaviour of materials with microstructure. This is due to the fact that the asymptotic solution is 
valid in a region close to the crack tip which is smaller than the characteristic lengths of the material, and thus it is of scarce physical relevance. A full field analysis is then 
required in order to grasp the qualitative and quantitative behaviour of the solution in a larger region and to be able to judge on the stress level supported by the material.

The static full field solution obtained by Radi (2008) for a Mode III crack by using Fourier transforms and Wiener--Hopf technique, shows that ahead of the crack tip within a zone smaller 
than the characteristic length in torsion, the total shear stress and reduced tractions occur with the opposite sign with respect to the classical LEFM solution, due to the relative 
rotation of the microstructural particles currently at the crack tip. However, this zone is found to have limited physical relevance and to become vanishing small for a characteristic 
length in torsion of zero. In this limit case, the solution recovers the classical $K_\modIII$ field with square root stress singularity. Outside the zone where the total shear stress is 
negative, the full field solution exhibits a bounded maximum for the total shear stress ahead of the crack tip, whose magnitude was adopted as a measure of the critical stress level 
for crack advancing. The introduction of a stress based fracture criterion thus defines a critical stress intensity factor, which increases with the characteristic length in torsion.

In this paper we analyse the steady-state propagation of a Mode III crack in couple stress elastic materials with finite characteristic lengths in bending and torsion. 
The static analysis is extended to the case of steady-state propagation in order to study the effects of inertia and crack-tip speed on the stress and deformation 
fields, as well as the variation of the fracture toughness due to the presence of microstructure. The paper starts with the description of the fully dynamical version of the couple 
stress elastic model in Section 2, followed by an overview of some peculiar dynamical effects that this model is able to simulate. Differently from conventional elastic theories, 
this reasonably simple model can indeed predict the dispersive character of shear wave propagation through elastic media with microstructures, for a specific range of values of the 
microstructural parameters. Results provided in Section 2.1 concerning dispersive wave propagation can thus be used to validate the characteristic length scales by comparison with 
atomistic and molecular dynamics simulations (Maranganti \& Sharma, 2007) or with experimental observations (Jakata \& Every, 2008). Next, the problem of steady-state crack 
propagation under Mode III loading conditions applied to the crack surfaces is presented in Section 2.2. The analytical full-field solution is then addressed in Section 3, making use of 
Fourier transform and Wiener--Hopf technique and basically following the approach introduced by Atkinson \& Leppington (1977) for a plane stationary crack problem. The crack tip is 
assumed to propagate at subsonic speed. Closed-form solutions are provided in Section 3.1 for the special case of vanishing rotational inertia, and in Section 3.2 for an arbitrary but 
small value of the rotational inertia parameter.

The stability of antiplane crack propagation is discussed in Section 4 by using the fracture criterion based on the maximum shear stress hypothesis introduced by Radi (2008), which can 
find application in crack propagation also. It follows, indeed, that propagation is unstable if the maximum shear stress occurring ahead of the crack tip during steady propagation is 
increasing as the crack tip speed increases. 

\section{Problem formulation}
\label{sec2}

Reference is made to a Cartesian coordinate system $(0, x_1, x_2, x_3)$ centred at the crack-tip at time $t = 0$. Under antiplane shear deformation, the
indeterminate theory of couple stress elasticity (Koiter, 1964) adopted here provides the following kinematical compatibility conditions
between the out-of-plane displacement $u_3$, rotation vector $\bvarphi$, strain tensor $\bepsilon$ and deformation curvature tensor $\bchi$
\beq
\label{comp1}
\epsilon_{13} = \frac{1}{2} \frac{\partial u_{3}}{\partial x_1}, \quad \epsilon_{23} = \frac{1}{2} \frac{\partial u_{3}}{\partial x_2}, \quad
\varphi_{1} = \frac{1}{2} \frac{\partial u_{3}}{\partial x_2}, \quad \varphi_{2} = -\frac{1}{2} \frac{\partial u_{3}}{\partial x_1},
\eeq
\beq
\label{comp2}
\chi_{11} = - \chi_{22} = \frac{1}{2} \frac{\partial^2 u_{3}}{\partial x_1 \partial x_2}, \quad
\chi_{21} = -\frac{1}{2} \frac{\partial^2 u_{3}}{\partial x_1^2}, \quad
\chi_{12} = \frac{1}{2} \frac{\partial^2 u_{3}}{\partial x_2^2}.
\eeq

Therefore, rotations are derived from displacements and the tensor field $\bchi$ turns out to be irrotational. According to the couple stress theory,
the non-symmetric Cauchy stress tensor $\bt$ can be decomposed into a symmetric part $\bsigma$ and a skew-symmetric part $\btau$,
namely $\bt = \bsigma + \btau$. In addition, the couple stress tensor $\bmu$ is introduced as the work-conjugated quantity of $\bchi^T$. For
the antiplane problem within the couple stress theory $\bepsilon$, $\bsigma$, $\btau$, $\bchi$ and $\bmu$ are purely deviatoric tensors. The reduced
tractions vector $\bp$ and couple stress tractions vector $\bq$ are defined as
\beq
\label{reduced}
\bp = \bt^T \bn + \frac{1}{2} \nabla \mu_{nn} \times \bn, \quad \bq = \bmu^T \bn - \mu_{nn} \bn,
\eeq
respectively, where $\bn$ denotes the outward unit normal and $\mu_{nn} = \bn \cdot \bmu \bn$. The conditions of dynamic equilibrium of forces
and moments, neglecting body forces and body couples, write
\beq
\label{equi}
\frac{\partial \sigma_{13}}{\partial x_1} + \frac{\partial \sigma_{23}}{\partial x_2} + \frac{\partial \tau_{13}}{\partial x_1} +
\frac{\partial \tau_{23}}{\partial x_2} = \rho \ddot{u}_{3}, \quad
\frac{\partial \mu_{11}}{\partial x_1} + \frac{\partial \mu_{21}}{\partial x_2} + 2\tau_{23} = J \ddot{\varphi}_{1}, \quad
\frac{\partial \mu_{12}}{\partial x_1} + \frac{\partial \mu_{22}}{\partial x_2} - 2\tau_{13} = J \ddot{\varphi}_{2},
\eeq
where $\rho$ is the mass density and $J$ is the rotational inertia.

Within the context of small deformations theory, the total strain $\bepsilon$ and the deformation curvature $\bchi$ are connected to stress
and couple stress through the following isotropic constitutive relations
\beq
\label{const}
\bsigma = 2G \bepsilon, \quad \bmu = 2G\ell^2 (\bchi^T + \eta \bchi),
\eeq
where $G$ is the elastic shear modulus, $\ell$ and $\eta$ the couple stress parameters introduced by Koiter (1964),
with $-1 < \eta < 1$. Both material parameters $\ell$ and $\eta$ depend on the microstructure and can be connected to the material characteristic
lengths in bending and in torsion, namely
\beq
\label{lengths}
\ell_b = \ell/\sqrt{2}, \quad \ell_t = \ell \sqrt{1 + \eta}.
\eeq

Typical values of $\ell_b$ and $\ell_t$ for some classes of materials with microstructure can be found in Lakes (1986, 1995). The limit value of
$\eta = -1$ corresponds to a vanishing characteristic length in torsion, which is typical of polycrystalline metals. Moreover, from the
definitions (\ref{lengths}) it follows that $\ell_t = \ell_b$ for $\eta = -0.5$ and $\ell_t = \ell = \sqrt{2} \ell_b$ for $\eta = 0$.
The constitutive equations of the indeterminate couple stress theory do not define the skew-symmetric part $\btau$ of the total stress tensor $\bt$, which
instead is determined by the equilibrium equations (\ref{equi})$_{2,3}$. Constitutive equations (\ref{const}) together with compatibility
relations (\ref{comp1}) and (\ref{comp2}) give stresses and couple stresses in terms of the displacement $u_3$:
\beq
\label{sigma}
\sigma_{13} = G \frac{\partial u_{3}}{\partial x_1}, \quad \sigma_{23} = G \frac{\partial u_{3}}{\partial x_2},
\eeq
\beq
\label{mu}
\mu_{11} = -\mu_{22} = G\ell^2(1 + \eta) \frac{\partial^2 u_{3}}{\partial x_1 \partial x_2}, \quad
\mu_{21} = G\ell^2 \left(\frac{\partial^2 u_{3}}{\partial x_2^2} - \eta \frac{\partial u_{3}}{\partial x_1^2}\right), \quad
\mu_{12} = -G\ell^2 \left(\frac{\partial^2 u_{3}}{\partial x_1^2} - \eta \frac{\partial^2 u_{3}}{\partial x_2^2}\right).
\eeq

The introduction of (\ref{mu}) into (\ref{equi})$_{2,3}$ yields
\beq
\label{tau}
\tau_{13} = -\frac{G\ell^2}{2} \Delta \frac{\partial u_{3}}{\partial x_1} + \frac{J}{4} \frac{\partial \ddot{u}_{3}}{\partial x_1}, \quad
\tau_{23} = -\frac{G\ell^2}{2} \Delta \frac{\partial u_{3}}{\partial x_2} + \frac{J}{4} \frac{\partial \ddot{u}_{3}}{\partial x_2},
\eeq
where $\Delta$ denotes the Laplace operator. By means of (\ref{sigma}) and (\ref{tau}), the equation of motion (\ref{equi})$_1$ becomes
\beq
\label{motion}
G \Delta u_3 - \frac{G\ell^2}{2} \Delta^2 u_3 + \frac{J}{4} \Delta \ddot{u}_3 = \rho \ddot{u}_3.
\eeq

\subsection{Preliminary analysis on shear wave propagation}

In this section we analyse the propagation of shear waves in a couple stress elastic material. We assume a solution of the governing
equation (\ref{motion}) in form of a planar shear wave as follows
\beq
\label{wave}
u_3(x_1,x_2,t) = C e^{i(k \mbox{\scriptsize \boldmath $\bx$} \cdot \mbox{\scriptsize \boldmath $\bn$} - \omega t)},
\eeq
where $C$ is the amplitude, $k$ is the wave number and $\omega$ the radian frequency. A substitution of (\ref{wave}) into (\ref{motion}) yields the following
dispersive equation
\beq
\label{disp}
k^4 + \frac{2}{\ell^2}\left(1 - \frac{\omega^2}{\theta^2}\right) k^2 - \frac{2}{\ell^2}\frac{\omega^2}{c_s^2} = 0,
\eeq
where $c_s = \sqrt{G/\rho}$ is the shear wave speed for classical elastic material and $\theta = \sqrt{4G/J}$.
By solving (\ref{disp}), we get the two equivalent dispersive relations
\beq
k^2 = -\frac{1}{\ell^2} \left(1 - \frac{\omega^2 h^2}{c_s^2}\right) +
\frac{1}{\ell^2} \sqrt{\left(1 - \frac{\omega^2 h^2}{c_s^2}\right)^2 + \frac{2 \omega^2 \ell^2}{c_s^2}}, \quad
\omega^2 = \frac{c_s^2}{2} \frac{k^2(k^2\ell^2 + 2)}{k^2 h^2 + 1},
\eeq
where $h$ is a characteristic length given by $h = c_s/\theta$.
We obtain that the shear wave is dispersive and propagates with phase speed
\beq
\label{disp2}
\tilde c^2 = \frac{\omega^2}{k^2} = \frac{c_s^2}{2} \left[ 1 - \frac{\omega^2 h_0^2 \ell^2}{c_s^2} + \sqrt{\left(1 - \frac{\omega^2 h_0^2 \ell^2}{c_s^2}\right)^2 +
2\frac{\omega^2 \ell^2}{c_s^2}} \right] = \frac{c_s^2}{2} \frac{k^2\ell^2 + 2}{k^2\ell^2h_0^2 +1},
\eeq
where $h_0 = h/\ell$ is a dimensionless parameter. Note that the relation (\ref{disp2}) was found by Ottosen et al. (2000) for the special case of
vanishing rotational inertia, $h_0 = 0$.

\begin{figure}[!htcb]
\centering
\includegraphics[width=7cm]{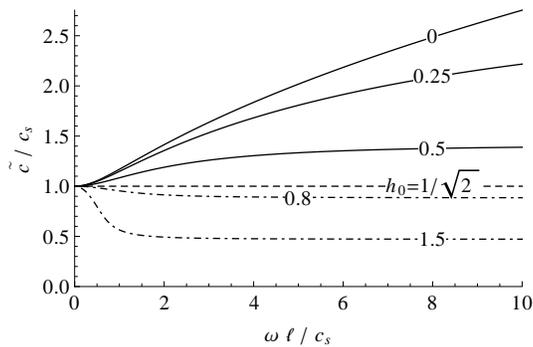}
\caption{\footnotesize Dispersive character of shear waves in couple-stress materials.}
\label{fig01new}
\end{figure}

The normalized shear wave speed $\tilde{c}/c_s$ as a function of the normalized frequency $\omega\ell/c_s$ is shown in Fig.~\ref{fig01new} for different
values of the parameter $h_0$, which is a measure of the rotational inertia. In the case of vanishing rotational inertia, $h_0 = 0$, the wave propagation is dispersive
with monotonically increasing phase speed, as found by Ottosen et al. (2000) (see also Askes \& Aifantis, 2011). In the presence of rotational inertia, $h_0 \neq 0$, the dispersive 
character of the
propagation is reduced until, for the limiting case $h_0 = 1/\sqrt{2}$, the wave propagation is non-dispersive as in classical elastic materials. Beyond this limit, the
wave propagation is again dispersive but with decreasing phase speed. It is worth noticing that the same result was found by Askes \& Aifantis (2011) in the framework of stable 
gradient elasticity. Similar behaviour was also reported by Georgiadis \& Velgaki (2003) for Rayleigh wave 
propagation in CS material. In particular, they analysed the behaviour of dispersive curves in the range $1/\sqrt{24} < h_0 < \sqrt{10/3}$ of the rotational inertia (or, 
in their 
notations, $0.25 < \hat{\mu} \hat{h}^2/\hat{\eta} < 20$, where $\hat{\mu} = G$, $\hat{\eta} = G\ell^2/2$ and $\hat{h}^2 = 3J/4\rho$). 
However, the authors of the aforementioned papers did not discuss
the limit value for the parameter $h_0$ separating the two essentially different regimes. 
Nevertheless, analysing the numerical results reported by Georgiadis and Velgaki (Figs. 5 and 6 in their paper), 
one could expect that the limit value coincides with our findings. Since this value is found for two different wave problems (antiplane shear waves in the present case and plane strain 
Rayleigh waves in their case), we believe that it may be a fundamental constant for CS materials. As it will be seen in Sec. \ref{sec-stability}, this limit value is also 
related to the stability of crack propagation.

%%%%%%%%%%%%%%%%%%%%%%%%%%%%%%%%%%%%%%%%%%%%%%%%%%%%%%%%%%%%%%%%%%%%%%
%%%%%%%%%%%%%%%%%%%%%%%%%%%%%%%%%%%%%%%%%%%%%%%%%%%%%%%%%%%%%%%%%%%%%%

It is also noted from Fig. \ref{fig01new} that, for low frequencies, $\omega \to 0$, the shear wave propagates with the same speed as in classical elastic materials, 
$\tilde{c} = c_s$, because long wavelengths do not feel the microstructure. Vice versa, for high frequencies, $\omega \to \infty$, short wavelengths interact with the 
microstructure and the shear wave propagates with constant speed depending on the rotational inertia, $\tilde{c} = c_s/\sqrt{2} h_0$. Consequently, in the range of 
rotational inertia $0 < h_0 \leq 1/\sqrt{2}$, the crack propagation is subsonic provided that the crack tip speed $V$ is less than $c_s$, whereas in the range 
$h_0 > 1/\sqrt{2}$, the crack propagation is subsonic subject to the condition $V < c_s/\sqrt{2} h_0$.

\subsection{Steady-state crack propagation}

We assume that the crack propagates with a constant velocity $V$ straight along the $x_1$-axis and is subjected to reduced force traction $p_3$
applied on the crack faces, moving with the same velocity $V$, whereas reduced couple traction $q_1$ is assumed to be zero,
\beq
\label{load}
p_3(x_1,0^\pm,t) = \mp \tau(x_1 - Vt), \quad q_1(x_1,0^\pm,t) = 0, \quad \text{for} \quad x_1 - Vt < 0.
\eeq
We also assume that the function $\tau$ decays at infinity sufficiently fast and it is bounded at the crack tip. These requirements
are the same requirements for tractions as in the classical theory of elasticity.

\begin{figure}[!htcb]
\centering
\includegraphics[width=5cm]{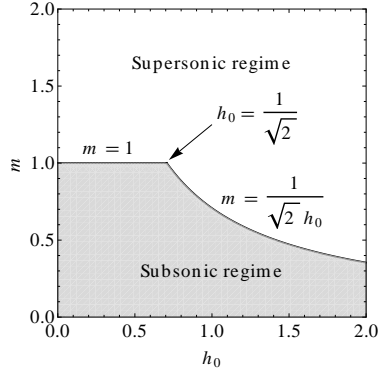}
\caption{\footnotesize Subsonic and supersonic regimes in the $m - h_0$ plane.}
\label{figsub}
\end{figure}

It is convenient to introduce a moving framework $X = x_1 - Vt$, $y = x_2$. By assuming that the out of plane displacement field has the form
\beq
\label{time}
u_3(x_1,x_2,t) = w(X,y),
\eeq
then the equation of motion (\ref{motion}) writes:
\beq
\label{pde}
\left(1 - m^2\right) \frac{\partial^2 w}{\partial X^2} + \frac{\partial^2 w}{\partial y^2} -
\frac{\ell^2}{2}\left(1 - 2m^2h_0^2\right)\frac{\partial^4 w}{\partial X^4} -
\ell^2\left(1 - m^2h_0^2\right)\frac{\partial^4 w}{\partial X^2 \partial y^2} - \frac{\ell^2}{2} \frac{\partial^4 w}{\partial y^4} = 0,
\eeq
where $m = V/c_s$ is the normalized crack velocity. We consider the subsonic regime (see Fig. \ref{figsub}), so that 
\beq
\label{min}
0 \leq m < \min \left\{ 1, \frac{1}{\sqrt{2} h_0} \right\}.
\eeq

%%%%%%%%%%%%%%%%%%%%%%%%%%%%%%%%%%%%%%%%%%%%%%%%%%%%%%%%%%%%%%%%%%%%%%
%%%%%%%%%%%%%%%%%%%%%%%%%%%%%%%%%%%%%%%%%%%%%%%%%%%%%%%%%%%%%%%%%%%%%%

According to (\ref{reduced}), the non-vanishing components of the reduced force traction and reduced couple traction vectors along the crack line $y = 0$, where
$\bn = (0, \pm1, 0)$, can be written as
\beq
\label{reduced2}
p_3 = t_{23} + \frac{1}{2} \frac{\partial \mu_{22}}{\partial X}, \quad q_1 = \mu_{21},
\eeq
respectively. By using (\ref{sigma})$_2$, (\ref{mu})$_{1,2}$, (\ref{tau})$_2$, and (\ref{reduced2}), the loading conditions (\ref{load}) on the upper crack
surface require the following conditions for the function $w$:
\beq
\label{bc1}
\frac{\partial w}{\partial y} - \frac{\ell^2}{2}\frac{\partial}{\partial y}\left[(2 + \eta - 2m^2h_0^2)\frac{\partial^2 w}{\partial X^2} +
\frac{\partial^2 w}{\partial y^2}\right] =
-\frac{1}{G} \tau(X), \quad
\frac{\partial^2 w}{\partial y^2} - \eta \frac{\partial^2 w}{\partial X^2} = 0, \quad \text{for} \quad X < 0, \quad y = 0^+.
\eeq
Ahead of the crack tip, the skew-symmetry of the Mode III crack problem requires
\beq
\label{bc2}
w = 0, \quad \frac{\partial^2 w}{\partial y^2} - \eta \frac{\partial^2 w}{\partial X^2} = 0, \quad \text{for} \quad X > 0, \quad y = 0^+.
\eeq

Note that the ratio $\eta$ enters the boundary conditions (\ref{bc1})-(\ref{bc2}), but it does not appear into the governing PDE (\ref{pde}).

\section{Full field solution}
\label{sec3}

In the present section the Wiener--Hopf analytic continuation technique (Noble, 1958; Freund, 1990) is used to obtain the full field solution
for a semi-infinite crack propagating in an infinite medium with constant velocity and subject to moving loading applied on the crack surfaces. Only the
upper half-plane ($y \geq 0$) is considered due to the skew-symmetry of the problem. Use of the Fourier transform and inverse transform is made.
For the function $w(X, y)$ they are
\beq
\label{fourier}
\overline{w}(s,y) = \int_{-\infty}^{\infty} w(X,y) e^{i s X} dX, \quad
w(X,y) = \frac{1}{2\pi} \int_{\mL} \overline{w}(s,y) e^{-i s X} ds,
\eeq
respectively, where $s$ is a real variable and the line of integration $\mL$ will be defined later.

Introduction of (\ref{fourier})$_2$ into the governing equation (\ref{pde}), yields the following ODE for $\overline{w}(s,x_2)$
\beq
\label{ode}
\frac{\partial^4 \overline{w}}{\partial y^4} - \frac{2}{\ell^2}\left[\ell^2\left(1 - 2m^2h_0^2\right)s^2 + 1\right]
\frac{\partial^2 \overline{w}}{\partial y^2} +
\frac{1}{\ell^2}\left[\ell^2\left(1 - 2m^2h_0^2\right)s^4 + 2\left(1 - m^2\right)s^2\right] \overline{w} = 0.
\eeq
Eq. (\ref{ode}) admits as bounded solution in the upper half-plane
\beq
\label{w0}
\overline{w}(s,y) = C(s) e^{-\alpha(s\ell) y/\ell} + D(s) e^{-\beta(s\ell) y/\ell}, \quad \text{for} \quad y \geq 0,
\eeq
where
\beq
\alpha(z) = \sqrt{1 + (1 - h_0^2m^2)z^2 + \chi(z)}, \quad
\beta(z) = \sqrt{1 + (1 - h_0^2m^2)z^2 - \chi(z)},
\eeq
\beq
\chi(z) = \sqrt{1 + 2(1 - h_0^2)m^2z^2 + h_0^4m^4z^4}.
\eeq
The function $\chi(z)$ has branch points at $\pm i b_1$ and $\pm i b_2$, where
\beq
b_1 = \frac{\sqrt{1 - h_0^2 + \sqrt{1 - 2h_0^2}}}{h_0^2 m}, \quad b_2 = \frac{\sqrt{1 - h_0^2 - \sqrt{1 - 2h_0^2}}}{h_0^2 m}.
\eeq
In the case $h_0 < 1/\sqrt{2}$, the branch points are located along the imaginary axis and the branch cuts are chosen to run from $i b_1$ to
$i b_2$ and from $-i b_1$ to $-i b_2$. In the opposite case, $h_0 > 1/\sqrt{2}$, the branch points are symmetrically located in the four quadrants
(separeted from the real axis) and the branch cuts are chosen to run from $i b_{1,2}$ to $i\infty$ and from $-i b_{1,2}$ to $-i\infty$. In both
cases the branch of the function $\chi(z)$ is chosen such that the function is positive along the real axis.

The function $\beta(z)$ has a branch point at $z = 0$, whereas the points $\pm i c$, where
\beq
c = \sqrt{\frac{2(1 - m^2)}{1 - 2h_0^2m^2}},
\eeq
are branch points of either $\alpha(z)$ or $\beta(z)$ depending on the values of the parameters $m$ and $h_0$. Consequently, in view of (\ref{min}), 
$c > 0$ and 
the additional branch cuts for the functions $\alpha(z)$ and $\beta(z)$ can be chosen to be along the imaginary axis. Again, the branches are
chosen such that the functions are positive along the real axis.

The functions $C(s)$ and $D(s)$ can be determined by imposing the boundary conditions (\ref{bc1}) and (\ref{bc2}). Applying the Fourier transform to the
function $w(X,0)$ and keeping into account (\ref{bc2})$_1$ we obtain
\beq
\label{one}
\overline{w}(s,0) = \overline{w}^-(s),
\eeq
which is analytic in the lower half complex $s$-plane, $\Im s < 0$. The Fourier transform of the reduced force traction $p_3$ and reduced couple traction $q_1$
at $y = 0$ leads to
\beq
\label{two}
\frac{\partial \overline{w}}{\partial y}(s,0) - \frac{\ell^2}{2} \left[-(2 + \eta - 2m^2h_0^2)s^2\frac{\partial \overline{w}}{\partial y}(s,0) +
\frac{\partial^3 \overline{w}}{\partial y^3}(s,0)\right] = -\frac{\overline{\tau}^-(s)}{G} + \frac{\overline{p}_3^+(s)}{G},
\eeq
\beq
\label{three}
\frac{\partial^2 \overline{w}}{\partial y^2}(s,0) + \eta s^2 \overline{w}(s,0) = 0,
\eeq
where $\overline{p}_3^+(s)$ is the Fourier transform of the reduced traction ahead of the crack tip and is analytic in the upper half complex $s$-plane, $\Im s > 0$.

A substitution of (\ref{w0}) into (\ref{one}), (\ref{two}) and (\ref{three}) allows us to find the functions $C(s)$ and $D(s)$, in the form
\beq
\label{cs}
C(s) = - \frac{(\beta^2 + \eta s^2\ell^2) \overline{w}^-(s)}{\alpha^2 - \beta^2},
\eeq
\beq
\label{ds}
D(s) = \frac{(\alpha^2 + \eta s^2\ell^2) \overline{w}^-(s)}{\alpha^2 - \beta^2},
\eeq
and also the following functional equation of the Wiener--Hopf type
\beq
\label{wh}
\overline{p}_3^+(s) + \frac{G}{2\ell}f(s\ell)\overline{w}^-(s) = \overline{\tau}^-(s),
\eeq
where
\beq
f(z) = \frac{1}{\alpha + \beta}
\Big\{ \alpha\beta(\alpha^2 + \beta^2 + 2\eta z^2) + \alpha^2\beta^2 - \eta^2 z^4 \Big\}.
\eeq

In order to apply the Wiener--Hopf technique of analytic continuation, Eq. (\ref{wh}) needs to be factorized into the product of two functions
analytic in the upper and lower half planes, respectively. This is described, in the following sections, separately for the two cases of vanishing
($J = 0$) and non-vanishing ($J \neq 0$) rotational inertia.

\subsection{The case of vanishing rotational inertia $J = 0$}

In the case of vanishig rotational inertia, we have $J = 0$ and consequently $h_0 = 0$. The Wiener--Hopf equation (\ref{wh}) reduces to
\beq
\label{wh0}
\overline{p}_3^+(s) + \frac{G}{2\ell}f(s\ell)\overline{w}^-(s) = \overline{\tau}^-(s),
\eeq
where
\beq
f(z) =  \frac{1}{\alpha + \beta} \left\{ \alpha\beta(\alpha + \beta)^2 - (\alpha\beta - \eta z^2)^2 \right\},
\eeq
and
\beq
\alpha(z) = \sqrt{1 + z^2 + \chi(z)}, \quad
\beta(z) = \sqrt{1 + z^2 - \chi(z)}, \quad
\chi(z) = \sqrt{1 + 2m^2z^2}.
\eeq
The function $\chi(z)$ has branch points at $\pm ib$, where $b = 1/(\sqrt{2}m)$, and the branch cuts are chosen to run from $\pm ib$ to
$\pm i\infty$ along the imaginary axis.

The function $\beta(z)$ has a branch point at $z = 0$, whereas the points $\pm i c$, where
\beq
c = \sqrt{2(1 - m^2)},
\eeq
are branch points of either $\alpha(z)$ or $\beta(z)$ depending on the values of the parameter $m$. Consequently, the additional branch cuts for
the functions $\alpha(z)$ and $\beta(z)$ can be chosen to be along the imaginary axis.

It is noted that $f(z)$ does not have any pole in the entire complex plane and has the following asymptotics at zero and infinity along the real
line
\beq
f(z) = \frac{1}{2}(1 + \eta)(3 - \eta) |z|^3 + O(|z|^2), \quad z \to \pm\infty, \quad z \in \Reals,
\eeq
\beq
f(z) = 2\sqrt{1 - m^2} |z| + O(|z|^2), \quad z \to 0, \quad z \in \Reals.
\eeq
Moreover, the only zero of $f(z)$ along the real axis is at $z = 0$. Note that $\eta=-1$ is a special case as the behaviour of the function 
$f(z)$ at infinite differs and consequently the solution is different. We will not consider this limit case.

\subsubsection{Wiener--Hopf factorization}
\label{sec311}

In order to apply the Wiener--Hopf technique of analytic continuation, eq. (\ref{wh0}) needs to be factorized into the product of two functions analytic in the
upper and lower half-planes, respectively.

Instead of the function $f(z)$, we will factorize the function
\beq
\label{k0}
k(z) = \frac{f(z)}{\sqrt{z^2} \Psi(z)},
\eeq
where
\beq
\label{psi_1}
\Psi(z) = \frac{1}{2} (1 + \eta)(3 - \eta)z^2 + 2\sqrt{1 - m^2}.
\eeq

It is noted that, along the real axis, the function $k(z)$ has no zeros and has the following asymptotics at zero and infinity
\beq
k(z) = 1 + O(|z|), \quad z \to 0, \quad z \in \Reals.
\eeq
\beq
k(z) = 1 + O(|z|^{-1}), \quad z \to \pm\infty, \quad z \in \Reals,
\eeq
Moreover, the function $k(z)$ has two simple poles at the points $\pm i d$, where
\beq
d = 2\frac{\sqrt[4]{1 - m^2}}{\sqrt{(1 + \eta)(3 - \eta)}}.
\eeq
The function $k(z)$ is factorized as follows
\beq
\label{fac1}
k(z) = \frac{k^-(z)}{k^+(z)}, \quad z \in \Reals,
\eeq
where $k^+(z)$ and $k^-(z)$ are analytic in the upper and lower half-planes, respectively, and are defined as follows
\beq
k^+(z) = \left\{
\barr{l}
\ds
e^{R(z)} \quad \text{for} \quad \Im z > 0, \\[3mm]
\ds
\frac{e^{R(z)}}{\sqrt{k(z)}} \quad \text{for} \quad \Im z = 0,
\earr
\right.
\eeq
\beq
k^-(z) = \left\{
\barr{l}
\ds
e^{R(z)} \quad \text{for} \quad \Im z < 0, \\[3mm]
\ds
\sqrt{k(z)} e^{R(z)} \quad \text{for} \quad \Im z = 0,
\earr
\right.
\eeq
where
\beq
\label{erre}
R(z) = -\frac{1}{2\pi i} \int_{-\infty}^\infty \frac{\log k(t)}{t - z} dt = -\frac{z}{\pi i} \int_{0}^\infty \frac{\log k(t)}{t^2 - z^2} dt =
-\frac{z}{\pi i} \left[\int_{0}^{1} \frac{\log k(t)}{t^2 - z^2} dt + \int_{0}^{1} \frac{\log k(1/t)}{1 - z^2t^2} dt\right],
\eeq
where use is made of the condition $k(-z) = k(z)$. The integrals in (\ref{erre}) must be evaluated as the Cauchy principal value if $\Im z = 0$.

The functions $k^+(z)$ and $k^-(z)$ have the following asymptotics at zero and infinity along the real axis
\beq
k^\pm(z) = 1 - \frac{z}{2\pi i} \int_{-\infty}^\infty \frac{\log k(t)}{t^2} dt + O(z^{2}), \quad z \to 0, \quad z \in \Reals,
\eeq
\beq
k^\pm(z) = 1 + \frac{1}{2\pi i z} \int_{-\infty}^\infty \log k(t) dt + O(z^{-2}), \quad z \to \pm\infty, \quad z \in \Reals.
\eeq
In the next section, we will also need to factorize the function $\sqrt{z^2}$. This is accomplished by writing
\beq
\sqrt{z^2} = z_+^{1/2} z_-^{1/2},
\eeq
where the branch cut for $z_\pm^{1/2}$ ranges from $0$ to $\mp i\infty$.
The functions $z_+^{1/2}$ and $z_-^{1/2}$ take along the real line the limit values
\beq
z_+^{1/2} = \left\{
\barr{ll}
z^{1/2}, & z > 0, \\[3mm]
i (-z)^{1/2}, & z < 0,
\earr
\right.
\qquad
z_-^{1/2} = \left\{
\barr{ll}
z^{1/2}, & z > 0, \\[3mm]
-i (-z)^{1/2}, & z < 0.
\earr
\right.
\eeq

\subsubsection{Solution of the Wiener--Hopf equation}

By making use of (\ref{k0}), the Wiener--Hopf equation (\ref{wh0}) becomes
\beq
\label{wh0b}
\overline{p}_3^+(s) + \frac{G}{4l}\sqrt{s^2\ell^2}[(1 + \eta)(3 - \eta)s^2\ell^2 + 4\sqrt{1 - m^2}]k(s\ell)\overline{w}^-(s) =
\overline{\tau}^-(s).
\eeq
The term $\sqrt{s^2\ell^2}$ is factorized as follows
\beq
\label{fac2}
\sqrt{s^2\ell^2} = (s\ell)_+^{1/2} (s\ell)_-^{1/2},
\eeq
where the functions $(s\ell)_+^{1/2}$ and $(s\ell)_-^{1/2}$ are analytic in the upper and lower half-plane and are defined in Sec. \ref{sec311}.

By using (\ref{fac1}) and (\ref{fac2}), eq. (\ref{wh0b}) can be written as
\beq
\label{wh0c}
\frac{k^+(s\ell)\overline{p}_3^+(s)}{(s\ell)_+^{1/2}} + \frac{G}{4l}(s\ell)_-^{1/2} [(1 + \eta)(3 - \eta)s^2\ell^2 + 4\sqrt{1 - m^2}]
k^-(s\ell)\overline{w}^-(s) =
\frac{\overline{\tau}^-(s)k^+(s\ell)}{(s\ell)_+^{1/2}}.
\eeq
We assume the following form of the loading applied on the crack faces
\beq
\tau(X) = \frac{T_0}{L} e^{X/L}, \quad X <0,
\eeq
that is, after Fourier transform,
\beq
\overline{\tau}^-(s) = \frac{T_0}{1 + isL},
\eeq
so that eq. (\ref{wh0c}) becomes
\beq
\label{wh0c2}
\frac{k^+(s\ell)\overline{p}_3^+(s)}{(s\ell)_+^{1/2}} + \frac{G}{4\ell}(s\ell)_-^{1/2} [(1 + \eta)(3 - \eta)s^2\ell^2 + 4\sqrt{1 - m^2}]
k^-(s\ell)\overline{w}^-(s) =
\frac{T_0 k^+(s\ell)}{(s\ell)_+^{1/2}(1 + isL)}.
\eeq
The right-hand side of (\ref{wh0c2}) can be easily split in the sum of plus and minus functions, namely
\beq
\label{fac3}
\frac{T_0 k^+(s\ell)}{(s\ell)_+^{1/2}(1 + isL)} =
\frac{T_0}{1 + isL} \left[ \frac{k^+(s\ell)}{(s\ell)_+^{1/2}} - \Xi \right]
+ \frac{T_0 \Xi}{1 + isL},
\eeq
where
\beq
\Xi = \frac{k^+(i\ell/L)}{(i\ell/L)_+^{1/2}}.
\eeq
The first term  of the right-hand side is no longer singular at $s = i/L$, and thus it is a plus function, whereas the last term is a minus
function.

By using (\ref{fac3}) and (\ref{fac1}), eq. (\ref{wh0c2}) can be factorized as follows
\beq
\label{wh0d}
\barr{l}
\ds
\frac{k^+(s\ell)\overline{p}_3^+(s)}{(s\ell)_+^{1/2}}
- \frac{T_0}{1 + isL} \left[ \frac{k^+(s\ell)}{(s\ell)_+^{1/2}} - \Xi \right] = \\[5mm]
\ds
- \frac{G}{4\ell}(s\ell)_-^{1/2}[(1 + \eta)(3 - \eta)s^2\ell^2 + 4\sqrt{1 - m^2}]k^-(s\ell)\overline{w}^-(s)
+ \frac{T_0 \Xi}{1 + isL}.
\earr
\eeq
The left and right hand sides of (\ref{wh0d}) are analytic functions in the upper and lower half-planes, respectively, and thus define an
entire function on the $s$-plane. The Fourier transform of the reduced force traction ahead of the crack tip and the crack opening gives
$\overline{p}_3^+ \sim s^{1/2}$ and $\overline{w}^- \sim s^{-5/2}$ as $s \to \infty$.

Therefore both sides of (\ref{wh0d}) are bounded as $s \to \infty$ and must equal a constant $F$ in the entire $s$-plane, according to
the Liouville's theorem.

As a result, we obtain
\beq
\label{resp3}
\overline{p}_3^+(s) =
\frac{T_0}{1 + isL} - T_0 \Xi\frac{(s\ell)_+^{1/2}}{k^+(s\ell)}\frac{1 - F(1 + isL)}{1 + isL},
\eeq

\beq
\label{resw}
\overline{w}^-(s) =
\frac{2T_0\Xi\ell}{G}
\frac{1 - F(1 + isL)}{(s\ell)_-^{1/2}(1 + isL) \Psi(s\ell) k^-(s\ell)},
\eeq
where the function $\Psi$ is defined in (\ref{psi_1}).

The constant $F$ is determined by the condition that the displacement $w(X)$ is zero at the crack tip $X = 0$, that is
\beq
\int_{-\infty}^{\infty} \overline{w}^-(s) ds = 0,
\eeq
which leads to
\beq
F =
\frac{\ds \int_{-\infty}^{\infty} \frac{ds}{(1 + isL)(s\ell)_-^{1/2} \Psi(s\ell) k^-(s\ell)}}
{\ds \int_{-\infty}^{\infty} \frac{ds}{(s\ell)_-^{1/2} \Psi(s\ell) k^-(s\ell)}}.
\eeq

\subsubsection{Analytical representation of displacements, stresses and couple-stresses}
\label{analex}
The reduced force traction ahead of the crack tip $p_3(X)$ and the crack opening $w(X)$ can be obtained from (\ref{resp3}) and (\ref{resw}),
respectively, by inverse Fourier transform, according to (\ref{fourier})$_2$. Since the integrand does not have branch cuts along the real line,
the path of integration $\mL$ coincides with the real $s$-axis. Further, we introduce the change of variable $\xi = s\ell$, thus obtaining
\beq
w(X) = \frac{T_0 \Xi}{\pi G} \int_{-\infty}^{\infty}
\frac{1 - F(1 + i\xi L/\ell)}{\xi_-^{1/2}(1 + i\xi L/\ell) \Psi(\xi) k(\xi) k^+(\xi)}
e^{-iX\xi/\ell} d\xi, \quad X < 0,
\eeq
\beq
p_3(X) = -\frac{T_0 \Xi}{2\pi \ell} \int_{-\infty}^{\infty}
\frac{1 - F(1 + i\xi L/\ell)}{1 + i\xi L/\ell} \xi_+^{1/2} \frac{k(\xi)}{k^-(\xi)}
e^{-iX\xi/\ell} d\xi, \quad X > 0.
\eeq
The Fourier transform of stress (symmetric and skew-symmetric) and couple stress fields can be derived from (\ref{w0}), (\ref{cs}), (\ref{ds})
and (\ref{sigma})--(\ref{tau}), namely
\beq
\overline{\sigma}_{23}(s,0) = -2T_0 \Xi \frac{\alpha\beta - \eta s^2\ell^2}{\alpha + \beta}
\frac{1 - F(1 + isL)}{(s\ell)_-^{1/2}(1 + isL) \Psi(s\ell) k^-(s\ell)},
\eeq

\beq
\barr{ll}
\overline{\tau}_{23}(s,0) &
\ds = -T_0 \Xi \frac{1}{\alpha + \beta}
\Big\{(1 + \eta)s^4\ell^4 + s^2\ell^2[c^2 + 2\eta + (1 + \eta)\sqrt{s^2\ell^2}\sqrt{s^2\ell^2 + c^2}]\Big\} \times \\[3mm]
 &
\ds \times \frac{1 - F(1 + isL)}{(s\ell)_-^{1/2}(1 + isL) \Psi(sl) k^-(s\ell)},
\earr
\eeq

\beq
\overline{\mu}_{22}(s,0) = -2T_0 \Xi \ell(1 + \eta) (is\ell) \frac{\alpha\beta - \eta s^2\ell^2}{\alpha + \beta}
\frac{1 - F(1 + isL)}{(s\ell)_-^{1/2}(1 + isL) \Psi(s\ell) k^-(s\ell)},
\eeq
The inverse Fourier transform can be performed as explained above, thus obtaining for $X > 0$
\beq
\sigma_{23}(X,0) = -\frac{T_0 \Xi}{\pi \ell} \int_{-\infty}^{\infty}
\frac{\alpha(\xi)\beta(\xi) - \eta \xi^2}{\alpha(\xi) + \beta(\xi)}
\frac{1 - F(1 + i\xi L/\ell)}{\xi_-^{1/2}(1 + i\xi L/\ell) \Psi(\xi) k^-(\xi)}
e^{-iX\xi/\ell} d\xi,
\eeq

\beq
\barr{l}
\ds \tau_{23}(X,0) = -\frac{T_0 \Xi}{2 \pi \ell} \int_{-\infty}^{\infty}
\frac{1}{\alpha(\xi) + \beta(\xi)}
\Big\{(1 + \eta)\xi^4 + \xi^2[c^2 + 2\eta + (1 + \eta)\sqrt{\xi^2}\sqrt{\xi^2 + c^2}]\Big\} \times \\[3mm]
\hspace{40mm} \ds \times \frac{1 - F(1 + i\xi L/\ell)}{\xi_-^{1/2}(1 + i\xi L/\ell) \Psi(\xi) k^-(\xi)}
e^{-iX\xi/\ell} d\xi,
\earr
\eeq

\beq
\mu_{22}(X,0) = -\frac{iT_0 \Xi}{\pi} (1 + \eta) \int_{-\infty}^{\infty}
\xi \frac{\alpha(\xi)\beta(\xi) - \eta \xi^2}{\alpha(\xi) + \beta(\xi)}
\frac{1 - F(1 + i\xi L/\ell)}{\xi_-^{1/2}(1 + i\xi L/\ell) \Psi(\xi) k^-(\xi)}
e^{-iX\xi/\ell} d\xi.
\eeq

\subsubsection{Results}

\begin{figure}[!htcb]
\centering
\includegraphics[width=14cm]{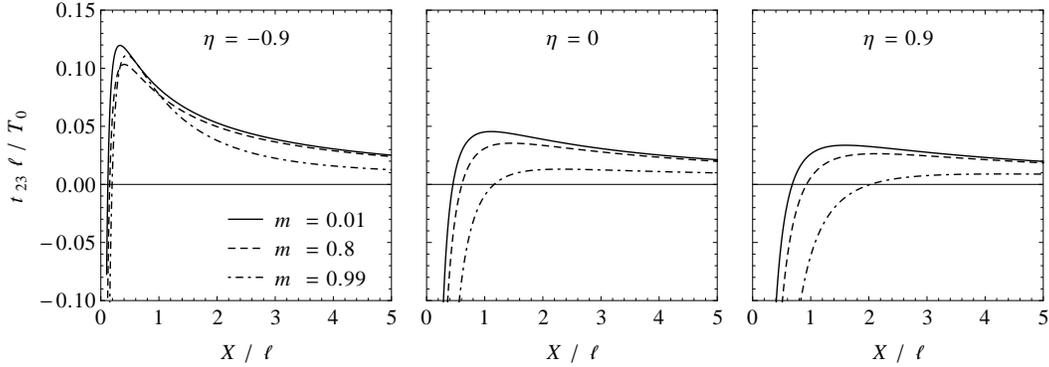}
\caption{\footnotesize Variation of total shear stress $t_{23}$ along the $X$-axis, ahead of the crack tip.}
\label{fig01}
\end{figure}

The variation of the normalized total shear stress ahead of the crack tip at $y = 0$, namely $t_{23} \ell / T_0$, versus normalized distance $X / \ell$ is shown in
Fig. \ref{fig01}, for three different values of $\eta = \{-0.9,0,0.9\}$ and three different values of the normalized crack tip speed $m = \{0.01,0.5,0.99\}$. The
steady-state solution approaches the static solution for small values of $m$ (see the solid lines in Fig. \ref{fig01} corresponding to $m = 0.01$), confirming earlier results
obtained by Radi (2008) for a stationary crack. In particular, the present results confirm that the strain-gradient effect is confined in a region of size $5\ell$ close to
the crack tip, where the solution gradually changes from the classical positive field, acting at large distance, to the negative near-tip asymptotic field.

%%%%%%%%%%%%%%%%%%%%%%%%%%%%%%%%%%%%%%%%%%%%%%%%%%%%%%%%%%%%%%%%%%%%%%
%%%%%%%%%%%%%%%%%%%%%%%%%%%%%%%%%%%%%%%%%%%%%%%%%%%%%%%%%%%%%%%%%%%%%%

As discussed in Radi (2008), the zone where the shear stress has the negative sign is very small and can be considered of no physical importance. Outside this zone, the
total shear stress attains a maximum, $t_{23}^\text{max}$, which is bounded and positive for $-1 < \eta < 1$ and $0 < m < 1$. As the crack tip speed increases, the total
shear stress tends to decrease ahead of the crack tip, and the rate of decrease is more pronounced for positive and large values of the ratio $\eta$, namely for large values
of the characteristic length in torsion according to (\ref{lengths})$_2$. Correspondingly, the zone ahead of the crack tip where the total shear has the negative sign enlarges 
and may reach the size of $2\ell$, where $\ell$ is proportional to the characteristic length in bending according to (\ref{lengths})$_1$.

\begin{figure}[!htcb]
\centering
\includegraphics[width=14cm]{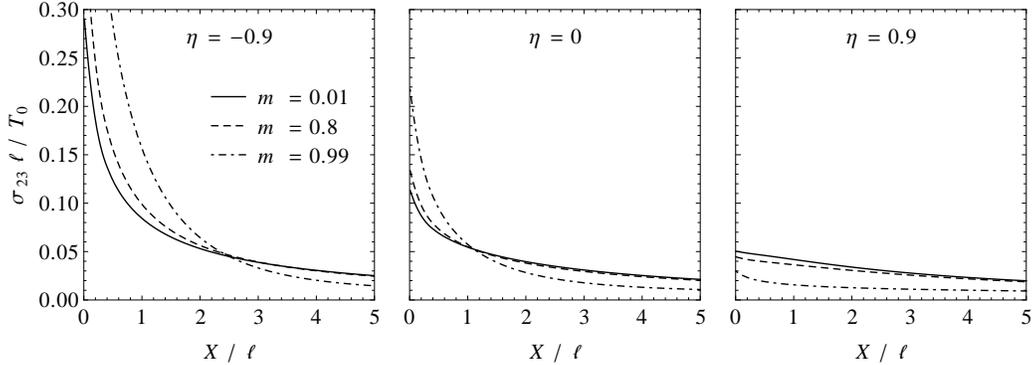}
\caption{\footnotesize Variation of symmetric shear stress $\sigma_{23}$ along the $X$-axis, ahead of the crack tip.}
\label{fig02}
\end{figure}

\begin{figure}[!htcb]
\centering
\includegraphics[width=14cm]{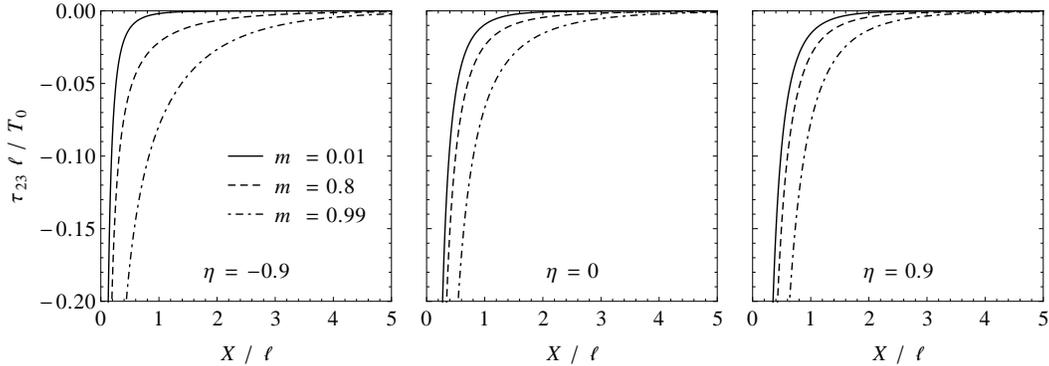}
\caption{\footnotesize Variation of skew-symmetric shear stress $\tau_{23}$ along the $X$-axis, ahead of the crack tip.}
\label{fig03}
\end{figure}

\begin{figure}[!htcb]
\centering
\includegraphics[width=14cm]{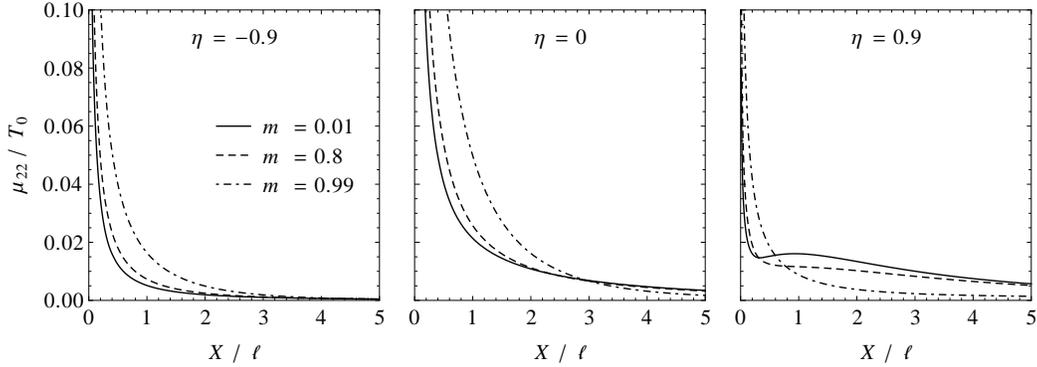}
\caption{\footnotesize Variation of the couple stress $\mu_{22}$ along the $X$-axis, ahead of the crack tip.}
\label{fig04}
\end{figure}

The corresponding normalized variations of the symmetric stress, skew-symmetric stress and couple stress are shown in Figs. \ref{fig02}, \ref{fig03} and \ref{fig04} for the
same values of $\eta$ and $m$. In Fig. \ref{fig02} it can be observed that the symmetric shear stress $\sigma_{23}$ is positive and finite ahead of the crack tip and tends to
display the singular behavior of classical elasticity as $\eta$ tends to $-1$. A different trend is observed as the crack tip speed becomes faster for negative or positive
values of $\eta$. Specifically, the symmetric stress increases with the crack speed for negative values of $\eta$, whereas it decreases for large and positive values of $\eta$.

%%%%%%%%%%%%%%%%%%%%%%%%%%%%%%%%%%%%%%%%%%%%%%%%%%%%%%%%%%%%%%%%%%%%%%
%%%%%%%%%%%%%%%%%%%%%%%%%%%%%%%%%%%%%%%%%%%%%%%%%%%%%%%%%%%%%%%%%%%%%%

The skew-symmetric shear stress $\tau_{23}$ ahead of the crack tip is negative and strongly singular (Fig. \ref{fig03}). Its magnitude increases with the crack tip speed for
every value of $\eta$, so that the total shear stress always decreases as the crack tip speed becomes faster, as already observed.

%%%%%%%%%%%%%%%%%%%%%%%%%%%%%%%%%%%%%%%%%%%%%%%%%%%%%%%%%%%%%%%%%%%%%%
%%%%%%%%%%%%%%%%%%%%%%%%%%%%%%%%%%%%%%%%%%%%%%%%%%%%%%%%%%%%%%%%%%%%%%

The plots in Fig. \ref{fig04} show that the couple stress $\mu_{22}$ decreases with $X$ and tends to vanish for $X \gg \ell$, thus recovering the classical solution of linear
elasticity. The decrease is very fast for negative values of the ratio $\eta$. For example, for $\eta = -0.99$ then $\mu_{22}$ completely vanishes at $X = 5\ell$,
whereas for positive values of $\eta$, namely for larger value of the characteristic length in torsion, the effects of couple stress and rotational gradients are felt at
larger distance from the crack tip. In particular, the non-monotonic behaviour of $\mu_{22}$ already observed in the static analysis (Radi, 2008) is recovered for very slow
crack tip speed, but it tends to disappear for faster speed of propagation.

%%%%%%%%%%%%%%%%%%%%%%%%%%%%%%%%%%%%%%%%%%%%%%%%%%%%%%%%%%%%%%%%%%%%%%
%%%%%%%%%%%%%%%%%%%%%%%%%%%%%%%%%%%%%%%%%%%%%%%%%%%%%%%%%%%%%%%%%%%%%%

\begin{figure}[!htcb]
\centering
\includegraphics[width=14cm]{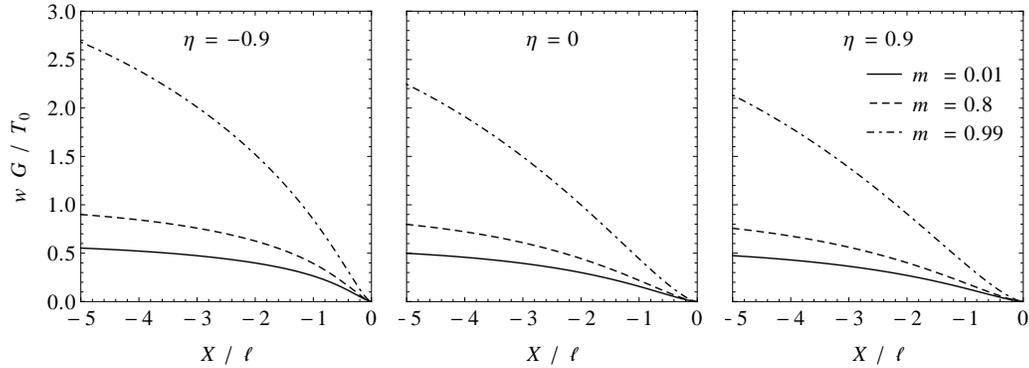}
\caption{\footnotesize Variation of the crack opening displacement $w$ along the crack face.}
\label{fig05}
\end{figure}

The normalized variations of the sliding displacement on the crack face, $w G /T_0$ at $y = 0$, versus normalized distance $X/\ell$ to the crack tip are shown in
Fig. \ref{fig05} for the selected range of values of $\eta$ and $m$. As well known, the crack tip profile is blunted for the classical elastic solution, but it
turns out to be sharp for CS elastic materials. The magnitude of the sliding displacement between the crack faces slightly decreases as $\eta$ increases from $-1$ to $1$,
thus indicating that the crack becomes stiffer due to the effects of the microstructure, here brought into play by strain rotational gradients. This trend is confirmed also
for crack propagation and proves that the microstructure may shield the crack tip from fracture, as already observed for stationary crack (Zhang et al., 1998; Georgiadis, 2003;
Radi, 2008). The results plotted in Fig. \ref{fig05} also show that the magnitude of the crack opening profile increases as the crack speed becomes faster, for every set
of material parameters. Since the crack tip sliding displacement is expected to be proportional to the square-root of the generalized J-integral (see Piccolroaz
et al., 2012), then an increasing
amount of energy is required in order to increase the crack tip speed, especially as the crack tip speed approaches the shear wave speed, thus predicting stable crack propagation
within the full subsonic regime.

%%%%%%%%%%%%%%%%%%%%%%%%%%%%%%%%%%%%%%%%%%%%%%%%%%%%%%%%%%%%%%%%%%%%%%
%%%%%%%%%%%%%%%%%%%%%%%%%%%%%%%%%%%%%%%%%%%%%%%%%%%%%%%%%%%%%%%%%%%%%%

The results of the full field analysis for very small values of $X$ plotted in Figs. \ref{fig01}--\ref{fig05} agree with the asymptotic solution obtained by Piccolroaz
et al. (2012). However, the radius of validity of the asymptotic solutions becomes vanishing small as $\eta$ approaches $-1$, namely as $\ell_t \to 0$. In this limit
case, the classical elastic solution is indeed recovered, since both the symmetric and the total shear stresses become unbounded at the crack tip and display the classical
square-root singularity.

\begin{figure}[!htcb]
\centering
\includegraphics[width=16cm]{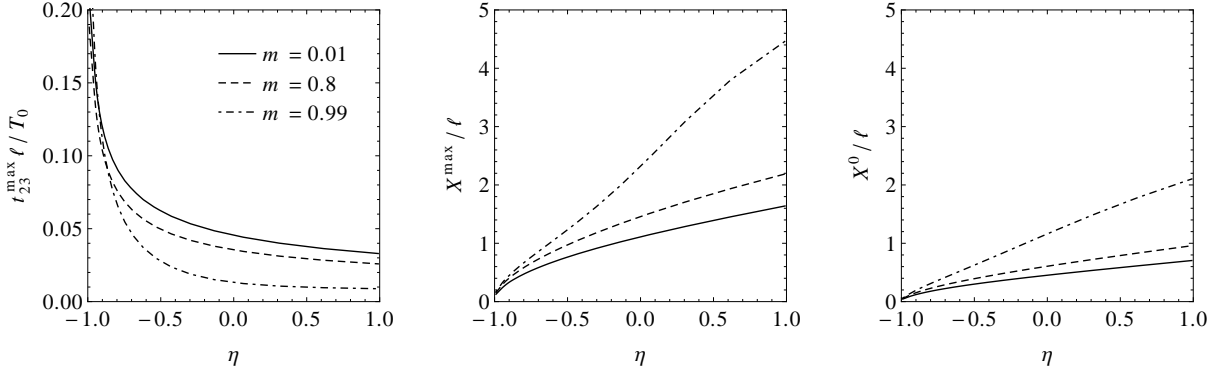}
\caption{\footnotesize Variations with the ratio $\eta$ of the maximum shear stress $t_{23}^\text{max}$ and its location $X^\text{max}$ and size $X^0$ of the
zone with negative shear stress ahead of the crack tip.}
\label{figtmax01}
\end{figure}

The variations of the non dimensional ratios $t_{23}\ell/T_0$, $X^\text{max}/\ell$ and $X^0/\ell$ with $\eta$ are plotted in Fig. \ref{figtmax01}. There, it can be observed
that the maximum shear stress decreases as $\eta$ increases, as well as the crack tip speed becomes faster. Correspondingly, both the distance to the crack tip of the
location of the maximum shear stress and the size of the zone ahead of the crack tip with negative shear stress become larger.

%%%%%%%%%%%%%%%%%%%%%%%%%%%%%%%%%%%%%%%%%%%%%%%%%%%%%%%%%%%%%%%%%%%%%%
%%%%%%%%%%%%%%%%%%%%%%%%%%%%%%%%%%%%%%%%%%%%%%%%%%%%%%%%%%%%%%%%%%%%%%

% \clearpage

\subsection{The case of non-vanishing rotational inertia $J \neq 0$}

\begin{figure}[!htcb]
\centering
\includegraphics[width=6cm]{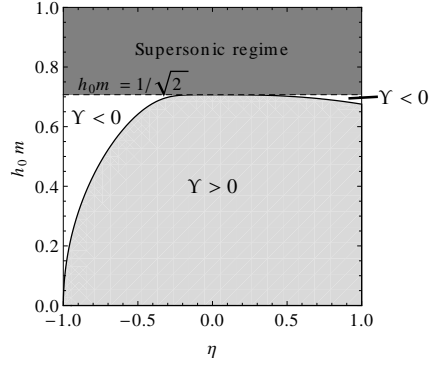}
\caption{\footnotesize Region of $\Upsilon > 0$ in the $\eta-h_0m$ plane.}
\label{figY}
\end{figure}

In the case $J \neq 0$, the asymptotics of the function $f(z)$ at zero and infinity along the real axis are as follows
\beq
f(z) = \Upsilon(\eta,h_0m) |z|^3 + O(|z|^2), \quad z \to \pm\infty, \quad
z \in \Reals,
\eeq
\beq
f(z) = 2\sqrt{1 - m^2} |z| + O(|z|^2), \quad z \to 0, \quad z \in \Reals,
\eeq
where
\beq
\Upsilon(\eta,h_0m) = \frac{1 - \eta^2 - 2h_0^2m^2 + 2\sqrt{1 - 2h_0^2m^2}(1 + \eta - h_0^2m^2)}{1 + \sqrt{1 - 2h_0^2m^2}}.
\eeq
Note that for $\eta = 0$ and any $h_0m < 1/\sqrt{2}$ this constant is always positive. However, if $\eta \neq 0$, there exists a set of parameters in the plane 
$(\eta,h_0m)$ where $\Upsilon(\eta,h_0m) \leq 0$ for all $h_0m < 1/\sqrt{2}$, see Fig. \ref{figY}. As for the case of vanishing rotational inertia $J=0$ and $\eta=-1$ 
described above, we are not considering the case when $\Upsilon(\eta,h_0m) \le 0$.

%%%%%%%%%%%%%%%%%%%%%%%%%%%%%%%%%%%%%%%%%%%%%%%%%%%%%%%%%%%%%%%%%%%%%%
%%%%%%%%%%%%%%%%%%%%%%%%%%%%%%%%%%%%%%%%%%%%%%%%%%%%%%%%%%%%%%%%%%%%%%

It is convenient to factorize the function $k(z)$ in the form (\ref{k0})
where $\Psi(z)$ is different and defined as follows:
\beq
\Psi(z) =\Upsilon(\eta,h_0m) z^2 + 2\sqrt{1 - m^2}.
\eeq
One can show that, under the condition $\Upsilon(\eta,h_0m) > 0$, the procedure for factorization is analogous to that reported in the previous section, and also the  
expressions of displacements, stresses and couple-stresses (except the skew-symmetric stress) can be represented in the same form as in Sec. \ref{analex}.

The Fourier transform of the skew-symmetric stress takes the form
\beq
\barr{ll}
\overline{\tau}_{23}(s,0) &
\ds = -T_0 \Xi \frac{1}{\alpha + \beta}
\Big\{\alpha^2\beta^2 + (\alpha^2 + \beta^2 + \alpha\beta)\eta s^2\ell^2 - (1 - 2h_0^2m^2)s^2\ell^2(\eta s^2\ell^2 - \alpha\beta)\Big\} \times \\[3mm]
 &
\ds \times \frac{1 - F(1 + isL)}{(s\ell)_-^{1/2}(1 + isL) \Psi(sl) k^-(s\ell)},
\earr
\eeq
which, after inversion, gives 
\beq
\barr{l}
\ds \tau_{23}(X,0) = -\frac{T_0 \Xi}{2 \pi \ell} \int_{-\infty}^{\infty}
\frac{1}{\alpha(\xi) + \beta(\xi)}
\Big\{\alpha^2(\xi)\beta^2(\xi) + (\alpha^2(\xi) + \beta^2(\xi) + \alpha(\xi)\beta(\xi))\eta \xi^2 - \\[3mm]
\ds \hspace{20mm}
(1 - 2h_0^2m^2)\xi^2(\eta \xi^2 - \alpha(\xi)\beta(\xi))\Big\} 
\frac{1 - F(1 + i\xi L/\ell)}{\xi_-^{1/2}(1 + i\xi L/\ell) \Psi(\xi) k^-(\xi)}
e^{-iX\xi/\ell} d\xi.
\earr
\eeq

\subsubsection{Results}

\begin{figure}[!htcb]
\centering
\includegraphics[width=14cm]{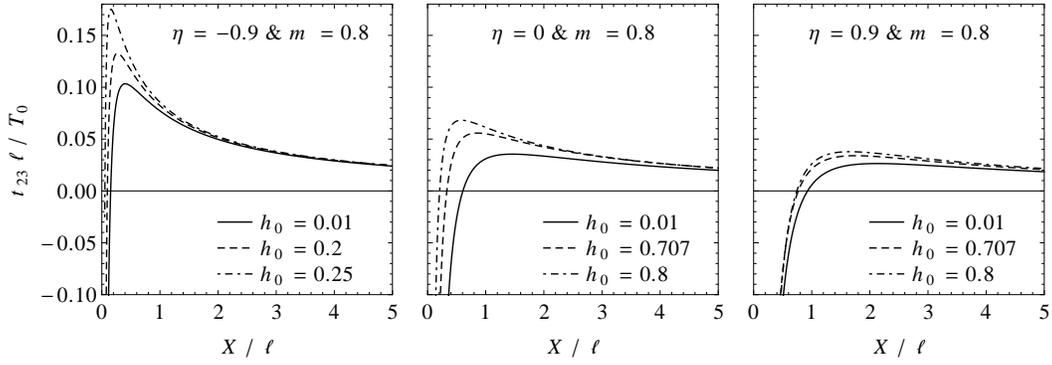}
\caption{\footnotesize Variation of total shear stress $t_{23}$ along the $X$-axis, ahead of the crack tip.}
\label{figj01}
\end{figure}

\begin{figure}[!htcb]
\centering
\includegraphics[width=14cm]{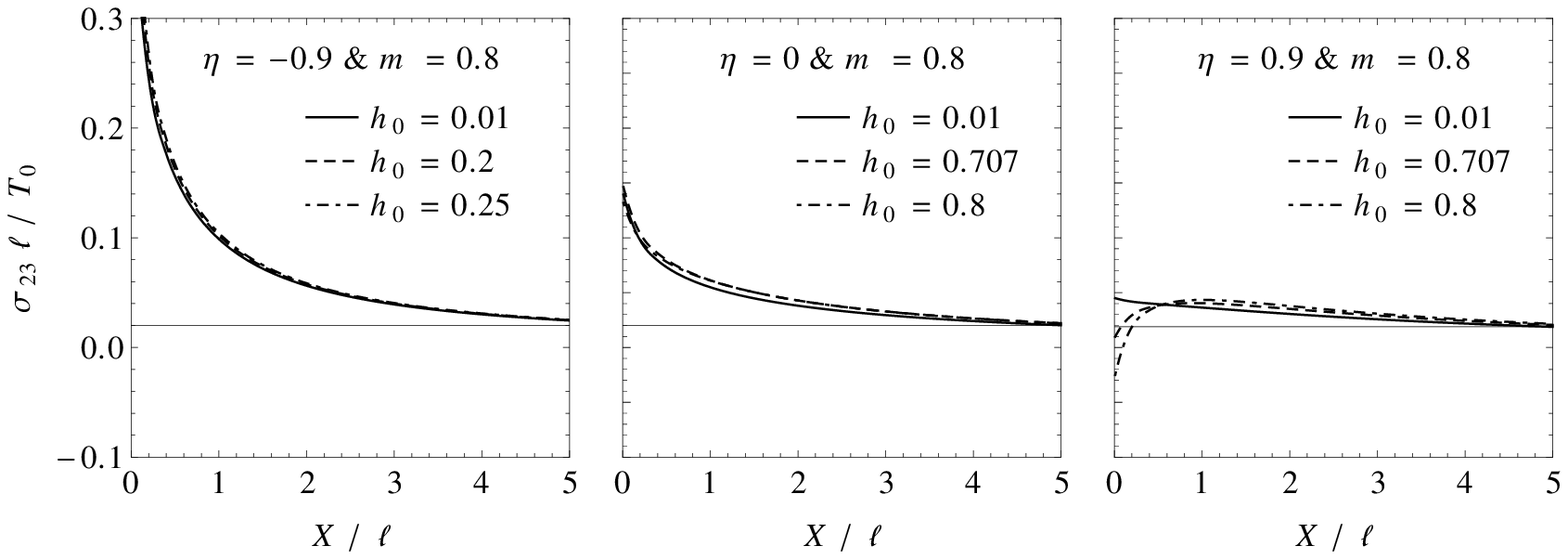}
\caption{\footnotesize Variation of symmetric shear stress $\sigma_{23}$ along the $X$-axis, ahead of the crack tip.}
\label{figj02}
\end{figure}

\begin{figure}[!htcb]
\centering
\includegraphics[width=14cm]{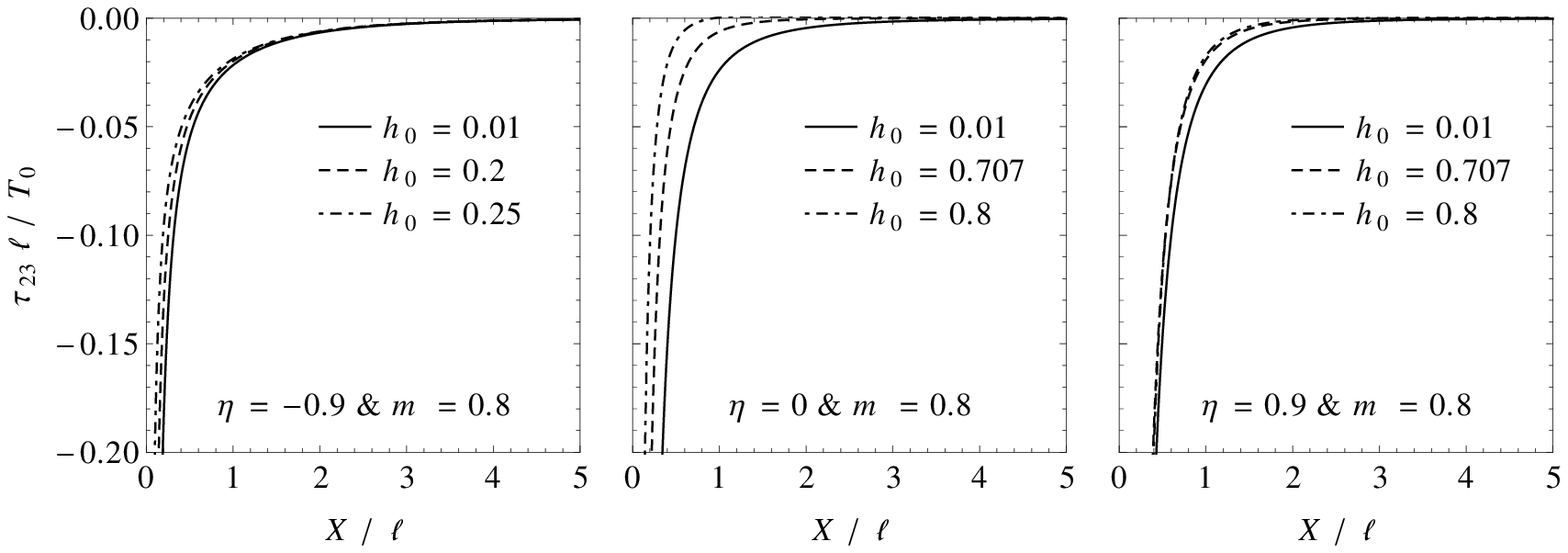}
\caption{\footnotesize Variation of skew-symmetric shear stress $\tau_{23}$ along the $X$-axis, ahead of the crack tip.}
\label{figj03}
\end{figure}

\begin{figure}[!htcb]
\centering
\includegraphics[width=14cm]{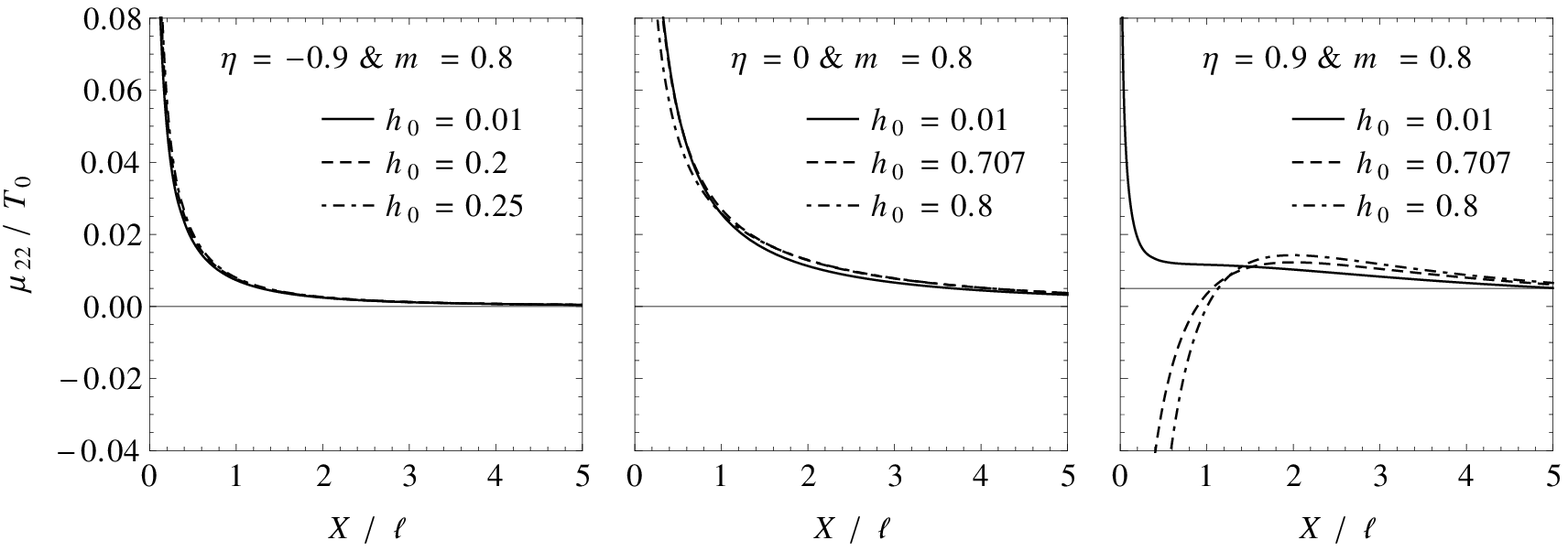}
\caption{\footnotesize Variation of the couple stress $\mu_{22}$ along the $X$-axis, ahead of the crack tip.}
\label{figj04}
\end{figure}

\begin{figure}[!htcb]
\centering
\includegraphics[width=14cm]{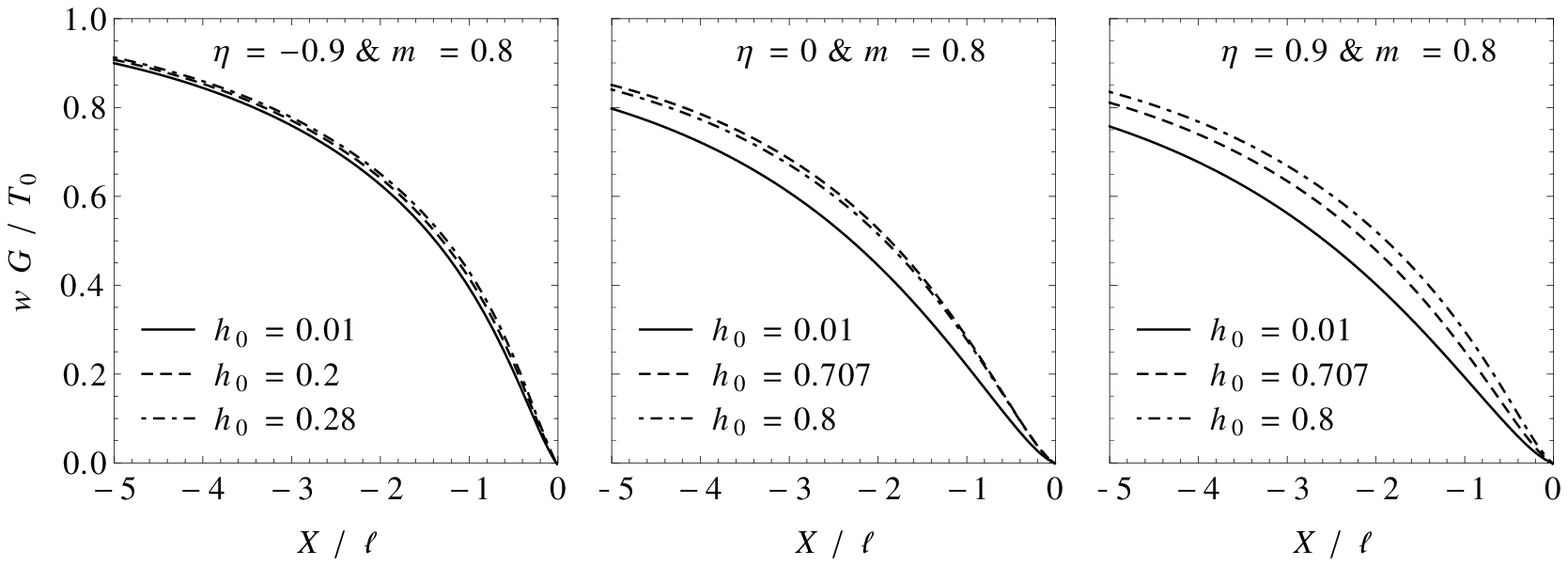}
\caption{\footnotesize Variation of the crack opening displacement $w$ along the crack face.}
\label{figj05}
\end{figure}

The effect of the rotational inertia on the stress, couple stress and displacement fields ahead of the crack tip can be observed in Figs. \ref{figj01}--\ref{figj05}.
There, the normalized variation of these fields are plotted for the same crack tip speed, namely for $m = 0.8$, for three different values of $\eta = \{-0.9, 0, 0.9\}$
and three different values of the non-dimensional parameter $h_0 = h/\ell$ providing for the contribution from rotational inertia. It can be observed that the effects
of rotational inertia on these fields mainly consist in the rising of the stress level ahead of the crack tip as the rotational inertia parameter $h_0$ is increased.

However, for large value of $\eta$ the symmetric shear stress display peculiar behavior just ahead of the crack tip, where it decreases and becomes negative for
$h_0 \ge 1/\sqrt{2}$. Correspondingly, the couple stress $\mu_{22}$ also change its sign ahead of the crack tip. As it will be discussed in the next section, such behavior
occurs as a consequence of the loss of stability of the crack propagation.

\begin{figure}[!htcb]
\centering
\includegraphics[width=16cm]{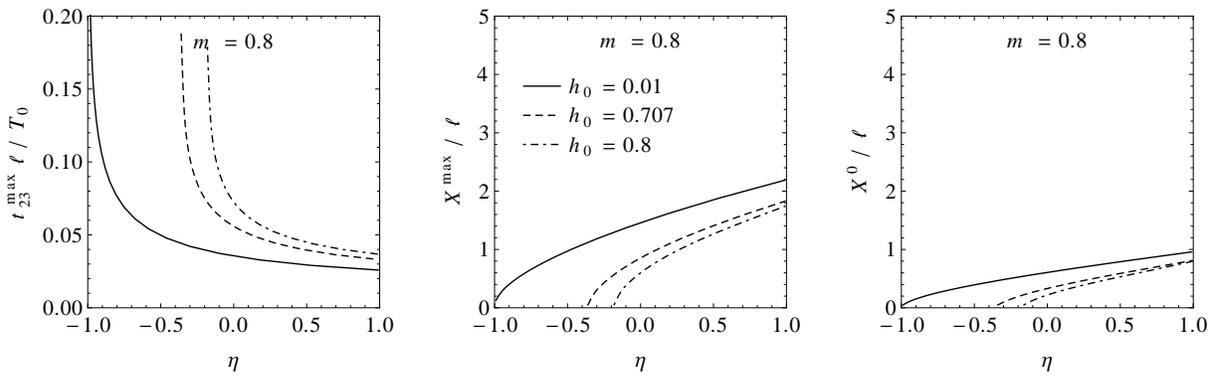}
\caption{\footnotesize Variations with the ratio $\eta$ of the maximum shear stress $t_{23}^\text{max}$ and its location $X^\text{max}$ and size $X^0$ of the
zone with negative shear stress ahead of the crack tip.}
\label{figjtmax01}
\end{figure}

Moreover, Fig. \ref{figj05} shows that the crack opening profile slightly increases with $h_0$ thus denoting that an additional amount of energy must be supplied from
the external loadings to allow the crack propagating at constant speed if the rotational inertia contribution is considered. Correspondingly, the maximum shear stress
increases its magnitude and tends to concentrate nearer to the crack tip, as illustrated in Fig. \ref{figjtmax01}.

%%%%%%%%%%%%%%%%%%%%%%%%%%%%%%%%%%%%%%%%%%%%%%%%%%%%%%%%%%%%%%%%%%%%%%
%%%%%%%%%%%%%%%%%%%%%%%%%%%%%%%%%%%%%%%%%%%%%%%%%%%%%%%%%%%%%%%%%%%%%%

%%%%%%%%%%%%%%%%%%%%%%%%%%%%%%%%%%%%%%%%%%%%%%%%%%%%%%%%%%%%%%%%%%%%%%
%%%%%%%%%%%%%%%%%%%%%%%%%%%%%%%%%%%%%%%%%%%%%%%%%%%%%%%%%%%%%%%%%%%%%%

%%%%%%%%%%%%%%%%%%%%%%%%%%%%%%%%%%%%%%%%%%%%%%%%%%%%%%%%%%%%%%%%%%%%%%
%%%%%%%%%%%%%%%%%%%%%%%%%%%%%%%%%%%%%%%%%%%%%%%%%%%%%%%%%%%%%%%%%%%%%%

%%%%%%%%%%%%%%%%%%%%%%%%%%%%%%%%%%%%%%%%%%%%%%%%%%%%%%%%%%%%%%%%%%%%%%
%%%%%%%%%%%%%%%%%%%%%%%%%%%%%%%%%%%%%%%%%%%%%%%%%%%%%%%%%%%%%%%%%%%%%%

%%%%%%%%%%%%%%%%%%%%%%%%%%%%%%%%%%%%%%%%%%%%%%%%%%%%%%%%%%%%%%%%%%%%%%
%%%%%%%%%%%%%%%%%%%%%%%%%%%%%%%%%%%%%%%%%%%%%%%%%%%%%%%%%%%%%%%%%%%%%%

%%%%%%%%%%%%%%%%%%%%%%%%%%%%%%%%%%%%%%%%%%%%%%%%%%%%%%%%%%%%%%%%%%%%%%
%%%%%%%%%%%%%%%%%%%%%%%%%%%%%%%%%%%%%%%%%%%%%%%%%%%%%%%%%%%%%%%%%%%%%%

\clearpage

\section{Stability analysis of the crack propagation}
\label{sec-stability}

In this section we discuss the stability of the crack propagation on the basis of the full field solution presented above. It must be pointed out that a sound criterion of 
fracture propagation in materials with microstructures cannot be based on the classical parameters of the crack-tip stress singularities. The singular behaviour occurs indeed at 
a distances to the crack tip smaller than the characteristic lengths, which can be commensurate with the microstructural size and even smaller. This would contradict the basic 
concept of continuum modelling which cannot predict phenomena occurring at the scales smaller than the microstructural size. According to that, Dyskin et al. (2010) recently proposed 
an intermediate asymptotic analysis for a finite length crack corresponding to the distances from the crack tip larger than the characteristic internal lengths but yet smaller than 
the crack length. 

In the present work, we adopt the fracture criterion of the maximum total shear stress, introduced in Radi (2008) and Geogiadis (2003), which usually occurs at 
distance commensurate with the microstructural size and, thus, still in the field of validity of continuum mechanics. The occurrence of a maximum positive value of the total shear 
stress ahead of the crack tip, $t_{23}^\text{max}$, allows us to formulate a simple criterion by assuming a critical shear stress level $\tau_C$ at which the crack starts propagating. 
The fracture criterion is then formulated as follows
\beq
\label{crit}
t_{23}^\text{max} = \tau_C,
\eeq
where $\tau_C$ is a material parameter.

\begin{figure}[!htcb]
\centering
\includegraphics[width=16cm]{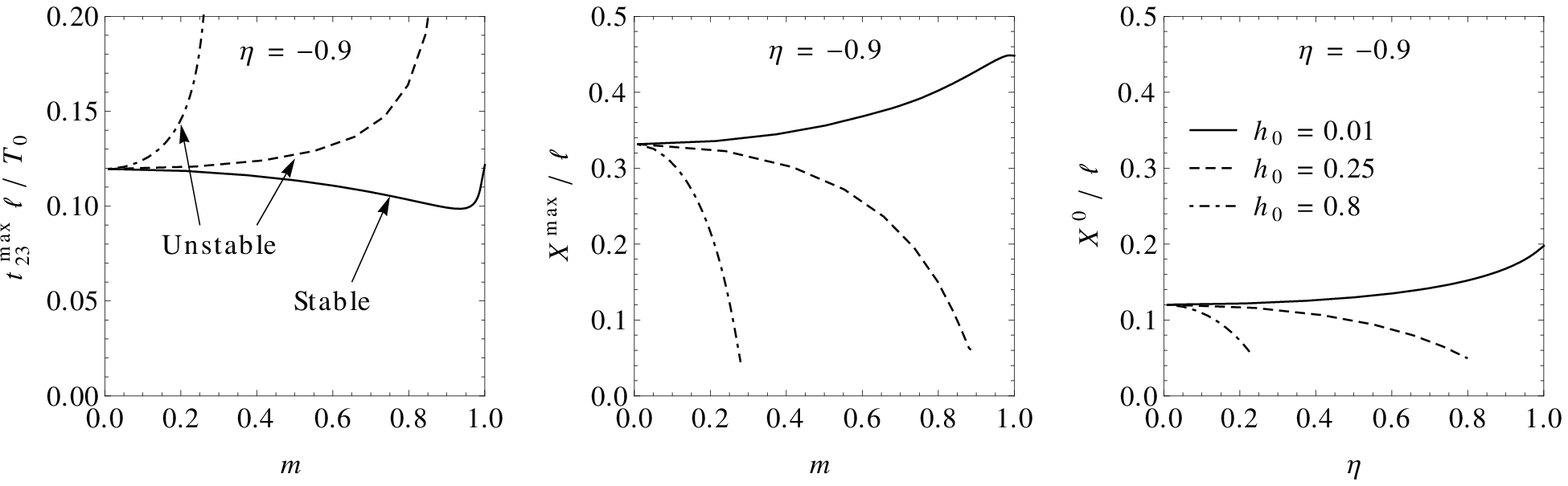}
\caption{\footnotesize Variations with the crack speed $m$ of the maximum shear stress $t_{23}^\text{max}$ and its location $X^\text{max}$ and size $X^0$ of the
zone with negative shear stress ahead of the crack tip.}
\label{figjtmax02.eta-09}
\end{figure}

\begin{figure}[!htcb]
\centering
\includegraphics[width=16cm]{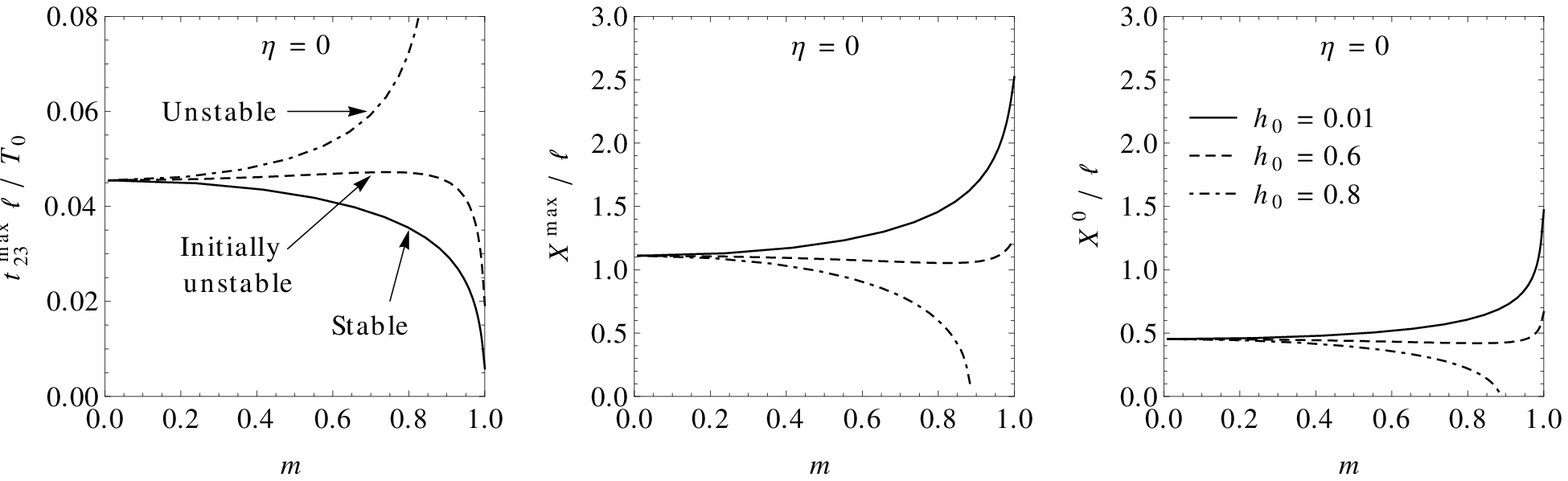}
\caption{\footnotesize Variations with the crack speed $m$ of the maximum shear stress $t_{23}^\text{max}$ and its location $X^\text{max}$ and size $X^0$ of the
zone with negative shear stress ahead of the crack tip.}
\label{figjtmax02.eta0}
\end{figure}

\begin{figure}[!htcb]
\centering
\includegraphics[width=16cm]{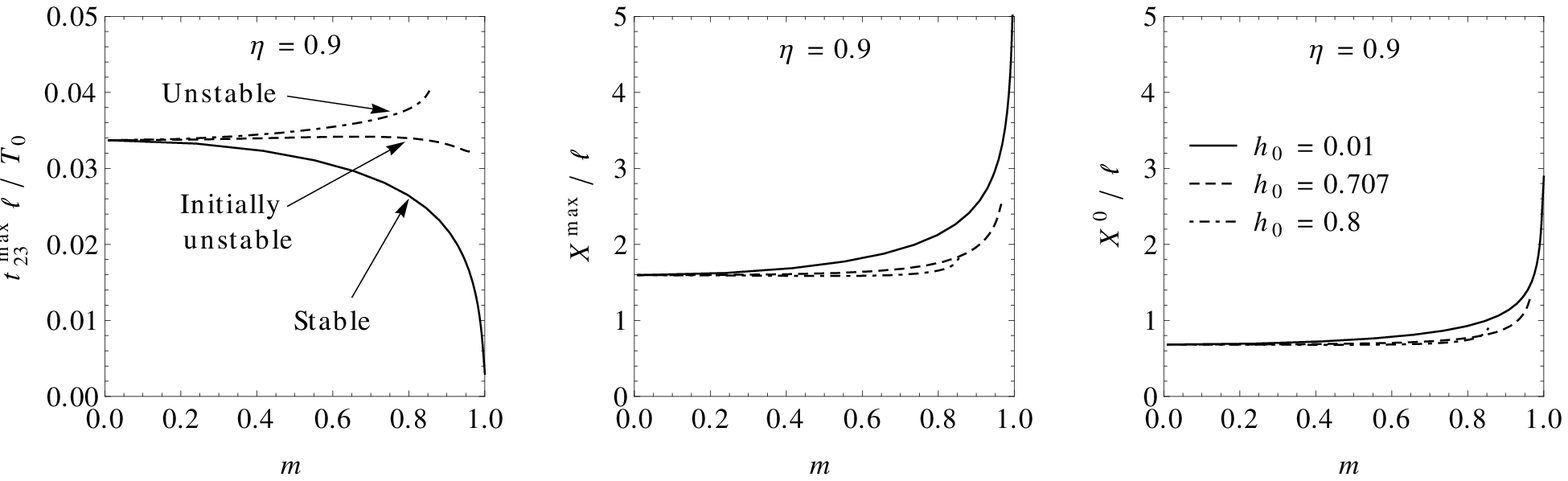}
\caption{\footnotesize Variations with the crack speed $m$ of the maximum shear stress $t_{23}^\text{max}$ and its location $X^\text{max}$ and size $X^0$ of the
zone with negative shear stress ahead of the crack tip.}
\label{figjtmax02.eta09}
\end{figure}

The normalized variations of the maximum attained by the total shear stress $t_{23}^\text{max}$, its location ahead of the crack tip and the size of the zone with
negative shear stress versus the normalized crack tip speed $m$ are plotted in Figs. \ref{figjtmax02.eta-09}, \ref{figjtmax02.eta0} and \ref{figjtmax02.eta09} for
$\eta = \{-0.9, 0, 0.9\}$, respectively. The results here reported show that for values of the rotational inertia parameter $h_0$ larger than $1/\sqrt{2}$ the
maximum shear stress $t_{23}^\text{max}$ grows very rapidly and becomes unbounded at a crack-tip speed lower than the shear wave speed $c_s$, namely for $m < 1$.
This behaviour necessarily implies a limit speed of propagation lower than $c_s$ and also unstable crack propagation if the stress based fracture criteria
(\ref{crit}) is assumed. Conversely, if the rotational inertia contribution is small, namely for $h_0 \ll 1/\sqrt{2}$, and $\eta$ is sufficiently greater than $-1$,
then the maximum shear stress decreases as the crack speed becomes faster and tends to zero as the crack speed approaches $c_s$ (solid lines in Figs.
\ref{figjtmax02.eta0}--\ref{figjtmax02.eta09}). Therefore, within this range of values for $\eta$ and $h_0$, the crack propagation turns out to be stable,
according to the fracture criteria (\ref{crit}). Non-monotonic behaviour is observed in the plot of $t_{23}^\text{max}$ versus $m$ for $\eta$ approaching the limit
value $-1$ (solid lines in Fig. \ref{figjtmax02.eta-09}, plotted for $\eta = -0.9$). In this case, the maximum shear stress first decrease as the crack tip speed
increases, thus suggesting the occurrence of stable crack propagation at speed sufficiently lower than the shear wave speed. However, as $m$ approaches $1$ then the
curve rapidly increases.

Therefore, the presence of microstructures provides a stabilizing effect on the propagation of crack. Indeed, for large values of the characteristic length in torsion, namely for 
large positive values of $\eta$, cracks can propagate steadily at high speed, also for large contribution of the rotational inertia. Differently, for small characteristic length 
in torsion, namely for negative values of eta, crack propagation becomes unstable as soon as the contribution of the rotational inertia becomes more pronounced. In particular, for 
vanishing small contribution from rotational inertia, crack propagation seems to be stable also for small size of the microstructures.

These findings in part agree with the results provided by Kennedy \& Kim (1993). These authors performed finite element analysis for the case of a crack propagating
at a constant velocity in a micropolar elastic material and observed that the micropolar effects consist in reducing the energy release rate for low velocity cracks
and became less pronounced with increasing crack propagation velocity.

%%%%%%%%%%%%%%%%%%%%%%%%%%%%%%%%%%%%%%%%%%%%%%%%%%%%%%%%%%%%%%%%%%%%%%
%%%%%%%%%%%%%%%%%%%%%%%%%%%%%%%%%%%%%%%%%%%%%%%%%%%%%%%%%%%%%%%%%%%%%%

%%%%%%%%%%%%%%%%%%%%%%%%%%%%%%%%%%%%%%%%%%%%%%%%%%%%%%%%%%%%%%%%%%%%%%
%%%%%%%%%%%%%%%%%%%%%%%%%%%%%%%%%%%%%%%%%%%%%%%%%%%%%%%%%%%%%%%%%%%%%%

%%%%%%%%%%%%%%%%%%%%%%%%%%%%%%%%%%%%%%%%%%%%%%%%%%%%%%%%%%%%%%%%%%%%%%
%%%%%%%%%%%%%%%%%%%%%%%%%%%%%%%%%%%%%%%%%%%%%%%%%%%%%%%%%%%%%%%%%%%%%%

\clearpage

\section{Conclusion}

The full field solution for a semi-infinite crack steadily propagating in a couple stress elastic material and subject to antiplane loading on the crack
surfaces has been obtained by using Fourier transform and the Wiener--Hopf technique. The material microstructure is characterized by finite characteristic
lengths in bending and torsion. The effect of rotational inertia is also considered in the analysis.

The main conclusion is that couple stress elasticity, providing for material characteristic lengths, may provide a clearer picture of the failure process near
a rapidly moving crack tip than classical elasticity. The analysis based on classical theory dictates that the only way a Mode III crack can respond to additional
influx of energy is by accelerating until it eventually outruns the energy at the shear wave speed. However, phenomena such as the actual limiting crack speed
under Mode III loading condition, which has been observed to be much lower than the shear wave speed as predicted by classical elastic theory, cannot at all be
addressed by classical, linear elastic theories. Experimental investigations (Livne et al., 2010) also bring to light the influence of the fracture process on
the limiting value. Since the process zone near the crack tip is strongly influenced by the microstructural parameters, such as the material characteristic lengths
and the rotational inertia, the maximum crack speed is also expected to be influenced by them. The fact that a fracture process zone has structure and dynamics
associated with its evolution provides for the dependence on microstructural parameters for the fracture energy and presents a possibility for explaining the
lower observed limiting speeds and the definition of the range of stable crack propagation. Recently, advanced models of the fracture process have been proposed in
literature, ranging from idealized atomistic and lattice models to phenomenological models to mechanism-based nucleation and growth models. While none of these
models have advanced to the stage of providing quantitative explanation of the limiting speed or for the dependence of the fracture energy on the crack speed, there
are qualitative features of the present micropolar model that appear promising.

\vspace{6mm}
{\bf Acknowledgements}. G.M. acknowledges the support from European Union FP7 project under contract number PIAP-GA-2011-286110. The paper has been completed during the Marie
Curie Fellowship of A.P. at Aberystwyth University supported by
the EU FP7 project under contract number PIEF-GA-2009-252857. E.R. gratefully acknowledges financial support from the ``Cassa di Risparmio di Modena'' in the
framework of the International Research Project 2009-2010 ``Modelling of crack propagation in complex materials''.

% \clearpage
% %%%%%%%%%%%%%%%%%%%%%%%%%%%%%%%%%%%%%%%%%%%%%%%%%%%%%%%%%%%%%%%%%%%%%%%
% %Appendix
% \appendix
% \renewcommand{\theequation}{\thesection.\arabic{equation}}
%
% \section{APPENDIX}
% \setcounter{equation}{0}
%
% \subsection{Appendix}
% \label{app01}

\end{document}